\documentstyle[12pt]{article}
\newcommand{\be}{\begin{equation}}
\newcommand{\ee}{\end{equation}}
\newcommand{\ba}{\begin{eqnarray}}
\newcommand{\ea}{\end{eqnarray}}

\newcommand{\no}{\nonumber\\}
\newcommand{\ci}[1]{\cite{#1}}
\newcommand{\bi}[1]{\bibitem{#1}}
\newcommand{\la}[1]{\label{#1}}
\def\gl#1{(\ref{#1})}

\newcommand{\dirak}{ /\!\!\!\!{\cal D}}
\newcommand{\leftvertex}{\widetilde{\mbox{\cal\bf M}}}
\newcommand{\rightvertex}{\widetilde {\mbox{\cal\bf M}}^{\dagger}}
\newcommand{\mcal}{\widetilde {\mbox{\cal\bf M}}}
\newcommand{\mdag}{\widetilde {\mbox{\cal\bf M}}^{\dagger}}
\newcommand{\phimone}{\phi_{m1}}
\newcommand{\phimtwo}{\phi_{m2}}
\newcommand{\philone}{\phi_{l1}}
\newcommand{\philtwo}{\phi_{l2}}
\newcommand{\Fi}[1]{f_{#1}( \tau )}
\newcommand{\Fio}[1]{f_{#1}( 0 )}
\newcommand{\fitm}{f_{t,m}\left(\frac{p^{2}}{\Lambda^{2}}\right)}
\newcommand{\fibm}{f_{b,m}\left(\frac{p^{2}}{\Lambda^{2}}\right)}
\newcommand{\fibl}{f_{b,l}\left(\frac{p^{2}}{\Lambda^{2}}\right)}
\newcommand{\fitl}{f_{t,l}\left(\frac{p^{2}}{\Lambda^{2}}\right)}
\newcommand{\fitmtau}{f_{t,m}(\tau)}
\newcommand{\fitltau}{f_{t,l}(\tau)}
\newcommand{\fibltau}{f_{b,l}(\tau)}
\newcommand{\fibmtau}{f_{b,m}(\tau)}
\newcommand{\fitmnull}{f_{t,m}(0)}
\newcommand{\fibmnull}{f_{b,m}(0)}
\newcommand{\fiblnull}{f_{b,l}(0)}
\newcommand{\fitlnull}{f_{t,l}(0)}
\newcommand{\dplus}{\Delta\left(\left(k+\frac{p}{2}\right)^{2}\right)}
\newcommand{\dminus}{\Delta\left(\left(k-\frac{p}{2}\right)^{2}\right)}
\newcommand{\xmt}{(x+m_{t}(k^{2}/\Lambda^{2}))}
\newcommand{\xmb}{(x+m_{b}(k^{2}/\Lambda^{2}))}
\newcommand{\ymt}{(y+m_{t}(k^{2}/\Lambda^{2}))}
\newcommand{\ymb}{(y+m_{b}(k^{2}/\Lambda^{2}))}
\newcommand{\kmt}{(k^{2}+m_{t}^{2}(k^{2}/\Lambda^{2}))}
\newcommand{\kmb}{(k^{2}+m_{b}^{2}(k^{2}/\Lambda^{2}))}
\newcommand{\mtk}[1]{m_{t}^{#1}(k^{2}/\Lambda^{2})}
\newcommand{\mbk}[1]{m_{b}^{#1}(k^{2}/\Lambda^{2})}
\newcommand{\deriv}{\left(\frac{k^{2}}{\Lambda^{2}}\right)}
\newcommand{\mesure}{k^{2}dk^{2}}
\newcommand{\lgt}{\ln\frac{\Lambda^2}{m_t^{2}}}
\newcommand{\lgb}{\ln\frac{\Lambda^2}{m_b^{2}}}
\newcommand{\mt}[1]{m_{t}^{#1}}
\newcommand{\mb}[1]{m_{b}^{#1}}
\newcommand{\lgtb}{\ln\frac{m_{t}^{2}}{m_{b}^{2}}}
\newcommand{\mesuretwo}{\frac{d\tau}{\tau}}
\newcommand{\remnant}{O\left(\frac{\ln
\frac{\Lambda^{2}}{\mu^2}}{\Lambda^{2}}\right)}
\newcommand{\ctone}{c_{t,1}}
\newcommand{\cttwo}{c_{t,2}}
\newcommand{\cbone}{c_{b,1}}
\newcommand{\cbtwo}{c_{b,2}}
\newcommand{\tr}{\mbox{tr}\!}
\newcommand{\dmu}{\partial_{\mu}}
\def\stick{\hbox to 0pt{\vbox to 7mm{}}}
\newcommand{\faza}[1]{e^{#1i\delta_{0}}}
\newcommand{\Lgg}[1]{\ln^{#1}\frac{\Lambda^2}{M_0^2}}
\newcommand{\lgg}[1]{\ln^{#1}\frac{\Lambda^2}{\mu^2}}
\newcommand{\acc}[1]{\left[
1+ O\left(\frac{1}{\ln^{#1}\frac{\Lambda^2}{\mu^2}}\right)\right]}
\newcommand{\nl}{\nonumber \\}
\newcommand{\mavr}{\langle\widetilde {\mbox{M}} \rangle}

\begin{document}
\renewcommand{\thefootnote}{\fnsymbol{footnote}}

\begin{center}
{\Large\bf
Two-Higgs Doublet Model from Quasilocal Quark Interactions}

\vspace{0.5cm}

{\bf A. A. Andrianov,
V. A. Andrianov,
V. L. Yudichev}\\
and\\
{\bf R. Rodenberg}\\
 {\sl Department of Theoretical Physics, \\
Institute of Physics, St.-Petersburg State University,\\
198904 St.-Petersburg, Russia}\\
$^b${\sl  III. Physikalisches Institut,\\
Abteilung f\"ur Theoretische Elementarteilchen Physik,\\
Physikzentrum, RWTH-Aachen,\\
D-52056,  Aachen, B.R.D.}
\bigskip
\end{center}

\begin{abstract}

Quark models with four-fermion interaction including derivatives of fields
in the strong coupling regime are used to implement
composite-Higgs extensions of the Standard
Model. In this approach the
dynamical breaking of chiral symmetry occurs in two (or more) channels
(near polycritical values for coupling constants),
giving rise to two (or more) composite Higgs doublets.
Two types of models are built for which Flavour Changing Neutral Currents
(FCNC)
are naturally suppressed. In the first Model I the second Higgs
doublet is regarded as a radial excitation of the first one. In the second
Model II the quasilocal Yukawa interaction with Higgs doublets reduces at low
energies to a conventional local one where each Higgs doublet
couples to a definite charge current and its v.e.v. brings the mass
either to up- or to
down-components of fermion doublets. For the
special configuration of four-fermion coupling constants
the dynamical CP-violation in the Higgs sector appears
as a result of complexity  of v.e..v. for Higgs doublets.
\end{abstract}

\newpage
\section{Introduction}
\hspace*{3ex}The fundamental particles of the Standard
Model (SM) of electroweak interactions, leptons, quarks
and gauge bosons, acquire masses through the interaction
with a scalar field (Higgs boson). The mass generation is mediated
by the Higgs mechanism which rests on the Electroweak
Symmetry Breaking (EWSB). To accommodate the well-
established electromagnetic and weak phenomena, the
Higgs mechanism requires the existence of at least one
iso-doublet, complex scalar field. After absorbing three
Goldstone modes to build massive states of  $W^{\pm}, Z$ bosons,
one degree of freedom remains, corresponding to
a real scalar particle (Higgs boson). If SM holds valid  as
a weak-coupling theory till very high energies then this particle cannot
have a mass heavier than few hundreds of $GeV$. Thus the search for
the Higgs boson is one of the fundamental quests for testing
the minimal SM. Current estimations  based on the different theoretical
requirements and experimental implications\cite{Gunion,Rod} give the
SM Higgs mass in the
"intermediate mass" window $65 < M_H < 200\, GeV$
for a top  quark mass value of about $175\, GeV$ \cite{CDF-D0}.
Despite of the recent successes of the SM in its excellent
agreement with the precision measurements at present
energies \cite{LEP}, it is generally believed that the SM
is not the final theory of elementary particle interactions.

There are many extensions of the SM
which lead to the enlargement of the Higgs sector of
the SM. For instance, the Minimal Supersymmetric
Standard Model (MSSM) \cite{Cab} entails two
elementary Higgs doublets at low energies, the
Two-Higgs-Doublet Model (2HDM) contains
two complex $SU(2)_L$ - doublet scalar fields with
hypercharge $Y=\pm1$ to couple the up-type /
down-type right-handed quarks to its Higgs doublet.
The search for relations between the many Higgs-field
dynamics and the masses of $t$-quark and Higgs boson
give the selection rule for a particular model beyond the
SM as well as for its acceptable parameters \cite{KrRod},
\cite{Kras}, \cite{ARR}, \cite{RK}. In more
complicated theories ( see \cite{Gunion} and references therein)
such as SUSY SM ones, $E_6 $ ones,
or Left-Right symmetric ones \cite{Moh}, several neutral scalars,
charged scalars and even double-charged scalars
are required in order to give all amplitudes acceptable
high-energy behavior \cite{Gunion,Rod}.

However there exists an alternative
possibility \cite{Bardeen} to restrict the number of
elementary particles to the observable fermion
and vector-boson sector with generation of scalar
Higgs particles due to attractive self-fermion
interaction. Namely, the quark self-interaction may
be responsible for the
production of quark-antiquark
bound states which are identified as composite
Higgs particles. The idea that the Higgs boson
could be a bound state of heavy quark pairs has
been developed and worked out in a series of
papers  \cite{Bardeen}, being
motivated by the  earlier work of Nambu and Jona-
Lasinio (NJL) \cite{NJL}.  In particular, for  $t$- quarks,
it is  provided by the Top-Mode Standard Model (TSM) Lagrangian,
known also as the Bardeen-Hill-Lindner (BHL) Lagrangian
\cite{Bardeen},\cite{LL},\cite{Lind}. The possibility that multiple
four-fermion interactions (for three and a heavy fourth
generations) are important in EWDSB, leading to
an effective 2HDM at low energies, has been
investigated in \cite{Luty}. In this model Higgs boson
induced Flavour Changing Neutral Currents (FCNS's)
are naturally suppressed \cite{GW}.
Some recent theoretical aspects and
questions of $\bar t t$- condensation frameworks one can
find in the review of \cite{Cvetic}.
In these scenarios the heavy top mass is
explained by the "top-condensation" where new strong
forces lead to the formation of  $\bar t t$ bound
states and the EWSB. In a minimal  version
of quark models the top-condensation was triggered
by a local four-fermion interaction.

The main goal of this paper is to give
the description of the design of the Quasilocal Quark
Models of type I and type II which provide two composite
Higgs doublets and in principle  satisfy
phenomenological restrictions
on FCNC suppression. The particular, sample choice
of formfactors and bare Yukawa coupling constants
is made to obtain estimations for typical mass spectra
in Model I and Model II.

We propose the quark models
 with {\sl Quasilocal}
four-fermion interaction \cite{4ferm} where the derivatives
of fermion fields are included into vertices to influence
on the formation of the second Higgs doublet. Such
extensions of the Higgs sector lead to a broad
spectrum of excited bound states, moreover they
may be viewed as more natural than other,  above
mentioned extensions since the particles involved
in EWSB
form only a ground state spectrum generic for SM. In these
quasilocal NJL-like quark models (QNJLM)
the symmetries do not forbid further higher
dimensional vertices and one should expect
that the ground states could be accompanied
by (radial) excitations with identical quantum
numbers but much higher masses \cite{2channel,Pallante95,Volkov1,Volkov2}.

Thus, from the
viewpoint of the 2HDM SM, the QNJLM
are attractive because:
i) it is an extension of the minimal TSM
which adds new phenomena (e.g. a broad
mass spectrum of bound states including
charged Higgs bosons);
ii) it is a minimal extension in that it adds the
fewest new arbitrary constants;
iii) it easily satisfies theoretical constraints on $\rho\,\simeq 1$
and the tree-level FCNC's suppression \cite{GW}
in accordance with the experimental evidence;
iv) such a Higgs structure is required in order to
build a model with the $CP$- violation \cite{Lee} because
the one-Higgs doublet interaction does not provide
any effect of dynamical $CP$- violation. We shall
show in a toy model with quasilocal four-fermion interaction
how $P$-parity breaks down
dynamically for the special choice of coupling
constants \cite{CP}.

This article is organized as follows:
Section 2 contains the simplest Gross-Neveu
model which reminds how the Dynamical Chiral
Symmetry Breaking ( DCSB) arises in the scalar
channel due to strong interaction. In Sec. 3, we
formulate the main rules for the construction of
the QNJLM which admits the polycritical regime.
Here the effective potential and the mass spectrum
for composite scalar and pseudoscalar states
 are derived for
them. For more evidence, in Subsec. 3.3, we investigate
the two-
channel  QNJLM in the large-log approximation.
 In the vicinity of  the tricritical
point all possible solutions
are analyzed. It turns out that there exist three phases
with different correlation lengths in the scalar channel.
Moreover the special phase of dynamical $P$-parity breaking
is found. In Sec.4 and 5
two types of models are built for which
FCNC suppression may be
naturally implemented. In Sec.4,
the first extension of the SM composite
two-Higgs bosons for QNJLM (2HQM) is
proposed where the second Higgs
doublet is regarded as a radial excitation of the first one.
The second
model is constructed in Sec.5 so that the quasilocal
Yukawa interaction with Higgs doublets reduces at low
energies to a conventional local Yukawa vertex where each Higgs doublet
couples to a definite charge current and its v.e.v. brings the mass
either to up-or to
down-components of fermion doublets.
In this version the top and bottom
masses are explained by "top-, bottom- condensations".
On the base of the effective potential for the Model II
and the positivity of its variation
the mass spectrum for composite states is
investigated. It is interesting that
for the special configuration of coupling
constants the dynamical $CP$-violation  appears in the
Higgs sector. In the Summary we discuss the obtained results
and a possibility to use them in different aspects of high energy physics.
The Appendix contains the calculation of the matrices of the second
variations for composite two-Higgs bosons in Model I ,II and the
effective potential of Model II for the special choice of quasilocal
formfactors.

\section{DCSB in Models with
Four-fermion Interaction and the Critical Regime}
\hspace*{3ex}Let us remind how the Dynamical Chiral Symmetry Breaking
(DCSB) arises in a model
with local 4-fermion interaction due to strong attraction in the scalar
channel. The simplest, Gross-Neveu (GN) model retaining the
scalar channel only can
be presented by the Lagrangian density,
in two forms (the Euclidean-space formulation is taken here),
\begin{equation}
{\cal L}=\bar q\not\!\! D q\,+\,\frac{g^{2}}{4N_{c}\Lambda^2}\,
(\bar q q)^{2}
\,=\,\bar q \bigl(\not\!\! D\,+\,i \phi (x)\bigr) q \,
+\frac{N_{c}\Lambda^2}{g^2}\,\phi^{2}(x), \la{GN}
\end{equation}
where $ \not\!\! D \,=\,i\gamma_{\mu}\partial_{\mu}$ and $q \equiv ( q_{i} )$
stands for color fermion f\/ields with $N_{c}$ components.
For the time being we take the
number of f\/lavours
$N_{F}=1$ and the current quark mass $m_{q}=0$. In Eq.\gl{GN}
the scalar auxiliary f\/ield $\phi(x)$
(a prototype of the Higgs field)  is introduced in order to describe
the dynamical symmetry breaking phenomenon in the large-$N_{c}$ limit.

This model is implemented by an $O(4)$-symmetric momentum cutof\/f $\Lambda$
for the fermion energy spectrum.
For a quark model the cutof\/f $\Lambda$ can be thought of as
 a separation scale which appears when evaluating the SM low-energy
ef\/fective  action from a more fundamental theory.
The regularized ef\/fective action $S_{eff}$ for auxiliary field,
\begin{equation}
Z^{\Lambda}(\phi)\,=\, \exp(- S_{eff})\,=\,\biggl\langle \exp\biggl(
-\int\! d^{4}x\,{\cal L} (\phi (x))\biggr)\biggr\rangle_
{\bar q q},
\end{equation}
possesses the mean-field extremum on constant
configurations $\phi = <\phi> \equiv m_{d} = const$.

The relevant ef\/fective
potential $V_{eff}$ can be obtained by integration over fermions,
\begin{equation}
V_{ef\/f} (\phi)\,=\, \frac{S_{eff}}{(vol.)} \,=
\,\frac{N_{c}}{8\pi^2}\left\{\frac{\Lambda^4}{2}\left({1\over 2} -
\ln\frac{\Lambda^2 + \phi^2}{\mu^2}\right)
 - \frac{\phi^2\Lambda^2}{2}
+ \frac{\phi^4}{2}\ln\frac{\Lambda^2 + \phi^2}{\phi^2} +
\frac{8\pi^{2}\Lambda^2\, \phi^2}{g^2}
\right\},
\end{equation}
where the constant $\mu$  is  a  normalization  scale  for  quark
fields.
Its extrema  can be derived  from  the  mass-gap  equation,
\begin{equation}
R(\phi)\,      \equiv\,\frac{4\pi^2}{N_{c}}\cdot
\frac{\partial V_{ef\/f}}{\partial \phi}
= \phi \,\left(\left(\frac{8\pi^2}{g^2}\, -\, 1\right)\Lambda^2 +\,\phi^2
\ln\frac{\Lambda^2 + \phi^2}{\phi^2}\right)\,=\,0 . \la{MG}
\end{equation}
The main contribution into Eq.\gl{MG} is given by a tadpole term in the
fermion loop which is related to a vacuum expectation value (v.e.v.)
of the scalar fermion density,
\begin{equation}
R( \phi ) = \,\phi \frac{8\pi^2 \Lambda^2}{g^2}\, +\, i\frac{4\pi^2}{N_{c}}
\langle \bar q q \rangle \, .
\end{equation}
The cutof\/f independence is realized with aid of fine-tuning,
$8\pi^2/g^2 \simeq 1 - O(1/ \Lambda^2)$. In the language of the theory of
critical phenomena it  is  equivalent  to  developing  our  model
around a critical or scaling  point where the quantum system undergoes the
second-order phase transition.  By  definition  the  critical
coupling  constant  is  $g_{crit}^2  =  8\pi^2$.  When  $g^2    <
g_{crit}^2$ the only solution of mass-gap Eq.\gl{MG} is $\phi =  0$,
while for $g^2 > g_{crit}^2$ there exists another  nontrivial
solution for dynamical mass $m_{d} \not= 0$ which brings the true minimum for
$V_{eff}$. Meanwhile the symmetric solution $\phi =  0$  does not
provide then a minimum anymore but realizes  a maximum.

The f\/ine-tuning states that the strong
 $\Lambda^2$-dependence should be compensated by the corresponding term
in the coupling constant,
\begin{equation}
\frac{8\pi^2}{g^2} = 1- \frac{m_0^2} {\Lambda^{2}} .
\end{equation}
Its practical meaning is evident, namely, one produces a  mass scale
for physical states which is much less than the cutoff scale governing
large radiative corrections.
The deviation scale $m_0^2 << \Lambda^2$ determines the
physical mass of scalar meson.
Indeed its kinetic term  can be obtained from the second
variation of $S_{eff}$ by
calculating the 1-fermion loop  diagram (see App. Fig.1),
\begin{eqnarray}
S_{eff}             &\,\simeq\,&
S_{eff}(\phi=m_{d})\,+\,\frac{1}{2}\int  \frac{d^4  p}{(2\pi)^4}\,
\tilde \phi (-p) \Gamma(p)\,\tilde \phi (p),\nonumber\\
\phi &\,=\,& m_{d}\,+\,\tilde \phi ,
\end{eqnarray}
where the inverse propagator of scalar field reads:
\ba
\Gamma (p) &=& \frac{2N_c\Lambda^2}{g^2} -N_c \int_{k<\Lambda}
\frac{d^4k}{(2\pi)^4}\mbox{\rm tr} \left[
( \not\! k + \not\! p/2 + im_{d})^{-1}
( \not\! k - \not\! p/2 + im_{d})^{-1}\right] \no
&=&  (m_\phi^2 + p^2) I (p^2) + O\left(\frac{1}{\Lambda^2}\right),
\ea
in the chirally invariant regularization of the fermion loop.
The  scalar  meson
mass is given by the remarkable  Nambu  relation  $m_\phi  \simeq  2
m_{d}$ and  the formfactor $I(p)$ is determined by the relation,
\be
I (p)= 2N_c \int_{k < \Lambda} \frac{d^4k}{(2\pi)^4}\frac{1}{(k +
\frac12 p)^2 + m_{d}^2}\frac{1}{(k - \frac12 p)^2 + m_{d}^2}.\la{l4}
\ee
In order that the  physical
mass parameters were insensitive to $\Lambda$, i.e.  $\partial_{\Lambda}
m_{d} \,=\,0$, the  scale  $m_0$  should  be   weakly
dependent,     $m_0^2\,\sim\,        m_{d}^{2}\,\ln({\Lambda^2}/
m_{d}^2)$, on the cutof\/f $\Lambda$.

What have we learned from the above model?
\begin{enumerate}
\item[(i)] The cut-off theory can be used for processes involving
momenta $p$ much less than $\Lambda$ for the purposes of
discarding high-energy states from the theory.
\item[(ii)] The mass scale of meson states is assumed to be
much less than $\Lambda$ which is implemented in the vicinity of
critical values of coupling constants.
\item[(iii)] As a result of DCSB in these models only one type of
scalar mesons (i.e. eventually one Higgs doublet) is created
in the large-$N_c$ approach.
\item[(iv)] In such models the (radial) excitations of composite
meson states are
not present in the large-$N_c$ approach.
\end{enumerate}
Meantime the conventional
quark models with local four-fermion interaction may not
represent a consistent part
of the Beyond-Standard Model (BSM) ef\/fective action
and conceivably they shall be extended with inclusion of
higher dimensional vertices which are not forbidden by
symmetries and  induce the appearance of
a reach spectrum of excited composite meson states.
\section{Quasilocal Quark Models and Polycritical Regime}
\subsection{Dominant
higher-dimensional vertices in  DCSB phase}
\hspace*{3ex}In order to involve in the theory the effects of
the discarded states  at scales of order $\Lambda$ it
is needed  to adjust the existing couplings
constants in the Lagrangian and to add new, quasilocal,
non-renormalizable interactions (vertices). These vertices
are polynomial in the fields and derivatives of the fields
and only a finite number of interactions is required
when working to a particular order in $\frac{p}{\Lambda}$,
where $p$ is a typical momentum in whatever process
is under study.

We examine the DCSB patterns in the mean-f\/ield approach
(large-$N_{c}$ limit)
and estimate the vertices with any number of fermion legs and derivatives.
The main rule to select out relevant vertices is derived from the
requirement of insensitivity in respect to the separation
scale $\Lambda$ following the conception of low-energy effective action
\cite{4ferm}.

We assume that:
\begin{itemize}
\item[(i)] $\Lambda^2$-order contributions from
different vertices are dominant in
creating the DCSB-critical
surface that is provided by cancellation of  all
contributions of $\Lambda^2$-order and defines the polycritical regime;
\item[(ii)] $\Lambda^0$-order contributions from vertices assemble in
the mean-f\/ield action
to supply fermions with dynamical mass
$m_{d}<<\Lambda$ which establishes the low-energy physical
scale;
\item[(iii)] respectively $\Lambda^{-2}$ (etc.)-order contributions
are irrelevant at energies much lower than $\Lambda$ and so
may be dropped from the theory if such accuracy is unnecessary.
\end{itemize}
In the large-$N_c$ approach the following approximation for v.e.v.
of fermion operators is valid,
\begin{equation}
\langle (\bar q q)^n \rangle \,=\, \biggl(\langle\bar q q\rangle
\biggr)^n \biggl( 1\,+\,O(1/N_c)\biggr),
\end{equation}
where any number of derivatives can be inserted between antifermion
and fermion operators.

V.e.v. of a bilinear operator is estimated in the assumption that quarks obtain
a dynamical mass. Namely,
\begin{equation}
\langle\bar q \biggl(\frac{\partial^{2}}{\Lambda^{2}}\biggr)^{n} q
\rangle\,  \sim  \,\frac{1}{\Lambda^{2n}}
\int \limits_{|p|<\Lambda} \frac{d^{4}p}{(2\pi)^{4}}\,\, tr \frac{p^{2n}}
{\not\! p + i m_d}\,\sim\, N_{c} m_d \Lambda^2 .
\end{equation}
One can see that the vertices with derivatives
in many-fermion interaction
are not suppressed and play equal role
in the mass-gap equation.

We omit the full classification of effective
vertices relevant in the mass-gap Eq. (see \ci{4ferm}) and report only the
minimal structure of the QNJLM which admits the polycritical regime,
\begin{equation}
{\cal L} = \bar q \not\!\! D q +  \frac{1}{N_{c}\Lambda^2}
\sum_{m,n = 0}^{l} a_{mn}\, \bar q_R
f_n \left(\frac{- \partial^2}{\Lambda^2} \right) q_L \cdot \bar q_L
f_m \biggl(\frac{- \partial^2}{\Lambda^2} \biggr) q_R,
\end{equation}
where $a_{mn}$ is a hermitian matrix of coupling constants
without zero eigenvalues and it is taken to be real symmetric one
in order that the interaction
did not break the  CP-parity explicitely.
Chiral fermion fields are given by $q_{L(R)} =1/2 (1\pm \gamma_5)
q$. We define the vertex formfactors to be polynomials of derivatives,
\be
f_m (\tau) = \sum^{K_m}_{i = 0} f^{(i)}_{m} \tau^{i},
\ee
to have quasilocal interactions. The variable $\tau$ is related to
derivatives, $\tau \rightarrow - \partial^2/\Lambda^2$.
We adopt the following rule for derivative action  which  provide
the hermiticity of fermion currents:
 \be
\bar q\frac{\partial^2}{\Lambda^2} q \equiv \frac14\bar q
\biggl(\frac{\stackrel{\rightarrow}{\partial}-
\stackrel{\leftarrow}{\partial}}{\Lambda}\biggr)^2 q. \label{4}
\ee
Besides let us regularize the interaction vertices with the help
of a momentum cutoff,
\be
\bar q q\longrightarrow\bar q\theta(\Lambda^2+\partial^2) q.
\label{5}
\ee
Without loss of generality one can choose formfactors $f_i(\tau)$ being
orthogonal polynomials on the unit interval,
\be
\int\limits_{0}^{1} d\tau
f_m(\tau) f_n(\tau)=\delta_{mn}. \label{6}
\ee
Let us now  introduce the appropriate set
of auxiliary f\/ields $\,\phi_{n}(x) \sim const$ and
develop the mean-f\/ield approach,
\begin{equation}
{\cal L}(\phi) \,=\, \bar q \biggl(\not\!\! D +
i M( \phi) P_L + i M^+( \phi) P_R\biggr) q
+ N_{c} \Lambda^{2} \sum_{m,n=1}^l \phi^{*}_{m}\,a_{mn}^{-1}\, \phi_{n}\, .
\end{equation}
The dynamical mass functional
is a linear combination of formfactors,
\begin{equation}
M(\phi) \equiv \sum_{n=1}^l
\phi_{n} (x) f_{n}\biggl(\frac{- \partial^2}{\Lambda^2}\biggr)\, .
\end{equation}
In accordance with  Eq.\gl{4} the differential
operator $M(\phi)$ is understood
as a Weyl ordered or fully antisymmetrized product of
functions $\phi_n$ and derivatives. Thereby we come to
a model with $l$  channels. When integrating out the fermion
fields one obtains the effective action of $\phi^*,\phi$ - fields.
The effective potential $V_{eff}$ is proved to be a functional depending
on the dynamical mass functional $M(\phi^*,\phi)$ and
proportional to $N_c$
that allows us to use the saddle point
approximation for $N_c >>1$.
\subsection{Effective potential  and equations on the mass spectrum
for QNJL Model}

The ef\/fective potential for auxiliary fields can be
derived with the  momentum  cutoff  regularization  by  averaging
over quark fields:
\ba
V_{eff}(\phi)&=&N_c\Lambda^2\sum_{m,n=1}^l \phi^{\star}_m a_{mn}^{-1}
\phi_n -  \frac{N_c\Lambda^4}{8\pi^2}\int\limits^1_0 d\tau
\tau \ln\left(1 + \frac{|M(\tau)|^2}{\Lambda^2\tau}\right)\no
&=&\frac{N_c}{8\pi^2}\biggl[\Lambda^2\sum_{m,n=1}^l
\phi^{\star}_m (8\pi^2 a_{mn}^{-1}
-\delta_{mn})\phi_n + \frac12|M_0|^4\biggl(\ln\frac{\Lambda^2}{|M_0|^2}+
\frac12\biggr)\no
&&+\frac12\int\limits_{0}^{1}\frac{d\tau}{\tau} (|M(\tau)|^4-|M_0|^4) +
O\biggl(\frac{1}{\Lambda^2}\biggr)\biggr],             \label{10}
\ea
herein $M_0\equiv M(0)$. The last approximation is valid in  such a
strong coupling regime where the dynamical mass $M_0 << \Lambda$. This
regime is of our main interest and it is realized in the vicinity of
a (poly)critical surface.  The critical values of coupling constants,
$a^c_{mn}=\delta_{mn}/ 8\pi^2$, are found from the cancellation of
quadratic divergences. In this paper we study the critical regime
in all $l$ channels. The vicinity of this polycritical point is described
by the following parametrization:
\be
8\pi^2 a_{mn}^{-1}\sim\delta_{mn}+\frac{\Delta_{mn}}{\Lambda^2}, \qquad
      |\Delta_{ij}|\ll \Lambda^2 .                           \label{11}
\ee
The generalized mass gap equations,
\be
\frac{\delta V_{eff}(\phi, \phi^*)}{\delta \phi_m^*} = 0 =
\frac{\delta V_{eff}(\phi, \phi^*)}{\delta \phi_m}, \label{9}
\ee
deliver the extremum to the effective
potential which may cause the DCSB if it is an absolute  minimum. They
read:
\ba
   \sum_{n=1}^l \Delta_{mn}\phi_n&=&
\int\limits_{0}^{1}\frac{d\tau}{\tau +
\frac{|M(\tau)|^2}{\Lambda^2}}(|M(\tau)|^2M(\tau) f_m(\tau)\no
&\simeq& f_m (0)|M_0|^2M_0\ln\frac{\Lambda^2}{M_0^2}\no
      &+&\int\limits_{0}^{1}\frac{d\tau}{\tau}(|M(\tau)|^2M(\tau) f_m(\tau)-
      |M_0|^2M_0 f_m (0)).                      \label{12}
\ea
It can be seen from the first relation that,
\be
   \sum_{m,n=1}^l \phi^{\star}_{m} \Delta_{mn}\phi_n =
\int\limits_{0}^{1}d\tau \frac{|M(\tau)|^4}{\tau +
\frac{|M(\tau)|^2}{\Lambda^2}} \geq 0,
\ee
which means that for the existence of a non-trivial dynamical mass
it is necessary to have at least one positive eigenvalue of the
matrix $\Delta_{mn}$. However not all the solutions provide
a  minimum  (see,  the  analysis  of  two-channel    models    in
\ci{CP}, \ci{2channel}).

The true minimum  is derived from the positivity of the second variation
of the effective action around a solution of the mass-gap equation,
\be
\phi_m = <\phi_m> + \sigma_m(x) + i \pi_m(x) .
\ee
This variation reads:
\ba
\frac{4\pi^2}{N_c} \delta^2 S_{eff} &&\equiv
\left(\sigma,(A^{\sigma\sigma} p^2 + B^{\sigma\sigma})
\sigma\right)\no
&&+ 2 \left(\pi,(A^{\pi\sigma} p^2 +
B^{\pi\sigma}) \sigma\right) + \left(\pi,(A^{\pi\pi} p^2 +
B^{\pi\pi}) \pi\right),
\ea
where two symmetric matrices - for
the kinetic term $\hat A = \left(A^{ij}_{mn}\right),\quad i,j =
(\sigma, \pi)$ and for the constant, momentum independent part,
$\hat B = \left(B^{ij}_{mn}\right)$ -  have been introduced.

The positivity of the second variation corresponds to the formation of
 physical mass spectrum for composite scalar and pseudoscalar states
which can be found from zeroes of the second variation determinant
at the Minkovski momenta ($p^2 < 0$),
\be
\det( \hat Ap^2+\hat B )=0.
\ee
Matrix elements of $\hat B$ are given  by the following relations:
\ba
B_{mn}^{\sigma\sigma}&=&
6\int\limits_0^1\frac{d\tau}{\tau}
\biggl[( \mbox{ Re}M)^2 f_m(\tau)f_n(\tau)
-M^2_0\Fio{m}\Fio{n}\biggr]\no
&+&M_0^2\Fio{m}\Fio{n}\left(6\Lgg{}-4\right) -2\Delta_{mn} \no
&+&2\int\limits_{0}^{1}\frac{d\tau}{\tau}(\mbox{ Im} M )^2\Fi{m}\Fi{n},
\label{32}  \\
B_{mn}^{\pi\pi}&=&
2\int\limits_0^1\frac{d\tau}{\tau}
\biggl[(\mbox{ Re}M)^2\Fi{m}\Fi{n}-M^2_0\Fio{m}\Fio{n}\biggr] \no
&+&2M_0^2\Fio{m}\Fio{n}\Lgg{}-2\Delta_{mn}\no
&+&6\int\limits_{0}^{1}\frac{d\tau}{\tau}(\mbox{ Im} M )^2\Fi{m}\Fi{n},
\label{33}
\ea
\be
B_{mn}^{\sigma\pi}=
4\int\limits_0^1\frac{d\tau}{\tau}( \mbox{Re}M )( \mbox{ Im}M )
\Fi{m}\Fi{n}, \label{34}
\ee
where the terms of $1/\Lambda^2$-order are neglected.

When exploiting the mass-gap equation
\gl{12} one can prove that the matrix $\hat B$ has always a zero
eigenvalue related to the eigenvector $ \phi^0_m=<\!\pi_m\!>\! - i\cdot
<\!\sigma_m\!> $.  It corresponds to the arising of the Goldstone  mode
(the massless Goldstone bosons).

 The kinetic energy matrix $\hat A$ turns out to be block-diagonal \ci{CP},
\be
A_{mn}^{\sigma\pi}=0,\quad A_{mn}^{\pi\pi}=A_{mn}^{\sigma\sigma}
\acc{},                                \label{38}
\ee
\begin{eqnarray}
  A_{mn}^{\sigma\sigma}&=&
\frac12\Biggl[\Fio{m}\Fio{n}\left( \Lgg{}+O\left( 1 \right) \right)\no
&+& \int\limits^{1}_{0}\left[\Fi{m}\Fi{n}-\Fio{m}\Fio{n} \right]
\frac{d\tau}{\tau}\Biggr]+O\left(\frac{1}{\Lambda^2} \right),
                                       \label{39}
\end{eqnarray}
herein we have displayed the leading terms only in the large-log
approximation. The more detailed expression can be found in \ci{CP},
\ci{2channel}.
\subsection{Quasilocal Two-Channel quark models
and possibility of dynamical
breaking  of P-parity}
For the further investigation of composite Higgs extensions of
the Standard Model
let us consider the Quasilocal Two-Channel quark model
in a tricritical point  \ci{CP}, \ci{2channel}.
We set $m,n=1,2$ in (12)-(20) and  retain only the lowest derivatives
in the potential, with
$f_1=1$,  $\, f_2=\sqrt{3}(1-2\tau)$. The dynamical
mass function is thereby,
$M(\phi)=\bar\phi_1+\bar\phi_2\sqrt{3}(1-2\tau)$.  As $\bar \phi_j$
are complex functions, $M(\phi)$ is complex too. However, with the global
chiral rotation $M(\phi)\rightarrow M(\phi)e^{i\omega}, \quad \omega=const$
it is always possible to implement $\mbox{Im}<M_0>_{\phi}=0$ and we
can choose the following parameterization:
\be
\bar\phi_1=\phi_1+i\rho, \quad
\bar\phi_2=\phi_2-i\frac{\rho}{\sqrt{3}},\quad \phi_i\equiv\mbox{ Re} \bar\phi_i .
\label{13}
\ee
The  equations~(\ref{9}) for the Two-Channel model read:
\ba
\Delta_{11}\phi_1+\Delta_{12}\phi_2&=&M_0^3\Lgg{} -6\sqrt{3}\phi_1^2\phi_2-
18\phi_1\phi_2^2-8\sqrt{3}\phi_2^3,\nl
d_1\phi_1-d_2\phi_2&=&2\sqrt{3}\phi_1(\phi_1^2+3\phi^2_2)+2\rho^2(
\frac{4}{\sqrt{3}}\phi_1-2\phi_2),\nl
\rho(\sqrt{3}\Delta_{11}-\Delta_{12})&=&2\rho\sqrt{3}(\phi_1^2+\phi_2^2+
\frac43\rho^2), \label{14}
\ea
where
\be
d_1=\sqrt{3}\Delta_{11}-\Delta_{12},\quad\quad
d_2=-\sqrt{3}\Delta_{21}+\Delta_{22} \label{15}.
\ee
We analyze the equations (\ref{14}) near a  polycritical point,
$|\Delta_{ij}|\sim {\mu}^2\ll \Lambda^2$, in the large-log
approximation ($\ln\frac{\Lambda^2}{{\mu}^2}\gg \ln\ln
\frac{\Lambda^2}{{\mu}^2}$). It gives rise to a set of
solutions.

For $\rho=0$ all the solutions are divided into the following classes:

  a) Gross-Neveu-like solutions $\phi_j^{GN}$ are:
\be
\phi^2_1=\frac{d_2^2\det
\Delta}{(\sqrt{3}d_1+d_2)^3\lgg{}}\acc{},\qquad
\phi_2\approx\frac{d_1}{d_2}\phi_1.
  \label{17}
\ee
These solutions deliver  minima to the potential when
$\sqrt{3}d_1+d_2 < 0$, with one  eigenvalue of the matrix $\Delta$
being in the over-critical regime and the other one in the sub-critical.

  b) Abnormal solutions are:
\be
\phi^2_1=\frac{\sqrt{3}d_1+d_2}{12}\acc{1/3}, \qquad
\phi_2\approx -\frac{\phi_1}{\sqrt{3}},  \label{19}
\ee
they correspond to the suppression of the large log-terms in
Eqs.(33) of motion and give  minima to the potential, when
$\sqrt{3}d_1+d_2 >0$, $\sqrt{3}d_1-2d_2\not =0$ ( either both
eigenvalues  of $\Delta$ are positive, or one is positive and
the other one is negative ).

c) On the planes $\sqrt{3}d_1+d_2=0$ and
$\sqrt{3}d_1-2d_2=0$ there appear special solutions with
different, peculiar asymptotics \ci{CP}, \ci{2channel}.

d) In general, in the models with more than one channel
complex solutions are allowed, and the imaginary parts of
all the variables $\phi_j$ cannot be removed simultaneously by
a global chiral rotation.  However the complex
solutions ($\rho\not=0$) minimize the effective action
only (!) for the narrow domain
in the vicinity of the plane
$\sqrt{3}d_1-2d_2=0$. Their asymptotic expressions are:
\be
\phi_1^2=\frac{d_1+4\Delta_{12}}{16\sqrt{3}(\lgg{}-3)} \qquad
\phi_2\approx -\sqrt{3}\phi_1,  \label{31}
\ee
and the dynamical mass is $m_c^2=4\phi_1^2$. The axial part of
the mass function looks as follows:

\be
\rho^2=\frac{d_1\sqrt{3}}{8}-\frac34( \phi^2_1+\phi^2_2 )=
\frac{d_1\sqrt{3}}{8}
\acc{}.                             \label{31'}
\ee

In each of the phase space domains  mentioned above one finds
four common boson states --- two scalar and two pseudoscalar
--- for real $\phi_j$, and, in general, --- for complex  $\phi_j$,
three states with mixed
P-parity and the pseudoscalar one with zero mass, the latest is in
accordance to the Goldstone theorem.

\hspace{5mm} The mass spectrum of related bosonic states
(collective excitations)
is determined by zero-modes of the matrix of second variations of
the ef\/fective
potential (25) and respectively by Eqs. (26) - (31).
Taking into account
the conditions necessary for a minimum of the potential,
 we find the solutions at $-m^2=p^2\le 0$,
giving physical values  of particle masses.

\hspace{5mm} In the case of $\rho=0$:

  a) NJL-like mass spectrum:
\ba
&& m_{\pi}^2=0\quad m_{\pi'}^2\approx m_{\sigma'}^2\approx -\frac{\sqrt{3}d_1+d_2}{3},\nl
&&m^2_{\sigma}\approx 4m_{d}^2,     \label{42}
\ea
 in this
domain  the radial excitation states are heavier than the lightest
scalar meson by a factor of logarithm.

  b) For the Abnormal solutions we have:
\ba
m_{\pi}^2=0,&& m_{\pi'}^2\approx \frac19\left( \frac43 \right)^{1/3}
\frac{( \sqrt{3}d_1-2d_2)^{4/3}}{( \sqrt{3}d_1+d_2 )^{1/3}}
\frac{1}{\lgg{1/3}},
\label{44} \\
m_{\sigma}^2\approx 6m_{d}^2,&&m_{\sigma'}\approx
\frac23( \sqrt{3}d_1+d_2 ).\nonumber
\ea
When comparing (39) and (\ref{44}) we find the scalar
channel correlation length to be different for each phase, that
corresponds to the tricritical point conditions.

c) For the
special real solutions the relations between scalar and
pseudoscalar meson masses are different
from (\ref{42}),(\ref{44}) (see \ci{CP},\ci{2channel}).

d)  Mass Spectrum in the P-parity Breaking Phase ( $\rho\not=0$).
One can see from (\ref{31}),(\ref{31'}) that in
the large-log approximation the axial dynamical mass
(the imaginary part of $M(\phi)$) dominates. It leads to
appearance of a massless boson in the scalar channel
in accordance to the Goldstone theorem. Conventionally, the
massless boson is related to be a pseudoscalar meson corresponding to
the generation of a real dynamical mass. In order to fit it
we make a global chiral rotation of fermionic fields $
q\rightarrow \exp( i\gamma_5\pi/4 )q $ accompanied by
corresponding rotation of the bosonic variables $\bar \phi_j
\rightarrow i\bar \phi_j $:
\be
\bar\phi_1=i\phi_1-\rho,\quad \bar\phi_2=i\phi_2+\frac{\rho}{\sqrt{3}}.
\label{57}
\ee
The classification of states given by the P-parity quantum
number is relevant only in the large-log approximation,
when:
\be
\frac{B^{\pi\sigma}}{B^{\sigma\sigma}}\approx
\frac{B^{\pi\sigma}}{B^{\pi\pi}}=O\left( \frac{1}{\lgg{}}
\right), \label{58}
\ee
next-to-leading logarithmic
effects are of no importance and one can neglect the mixing of
states with different P-parity.
Then the mass
spectrum of mesons is:
\begin{eqnarray}
  m_{1}^2=0,&& m_{2}^2\approx \frac{d_1+4\Delta_{12}}{\sqrt{3}\lgg{}}
\approx 16\phi_1^2=4m_d^2,
\label{59}      \nl
m_{3}^2\approx \sqrt{3}d_1,&& m_{4}^2\approx
\frac{4( d_1+\Delta_{12} )}{9\sqrt{3}\lgg{}}.
\end{eqnarray}
The ratio of $m_{2}$ and $m_{4}$
does not depend on the logarithm, so both the masses are comparable.
On the other hand, in the
models with a finite momentum cut-off,
when the effects of order of $1 / \lgg{}$
make sense,
 the dynamical P-parity
breaking is induced, since $ B_{\pi\sigma}\not=0$.
This phenomenon  of dynamical P-parity breaking can be used in
extensions of the Standard Model \cite{Bardeen} where several
Higgs bosons are composite ones.

Thus we conclude that the models with polycritical (tricritical)
points are
drastically different from the local NJL models in the variety
of the physical phenomena in the DCSB. Explorations of such
QNJLM in extensions of the SM
are pretty well motivated as the
underlying dynamics responsible for the top quark condensate
should most likely lead to a broad spectrum of excited
states, just like the hadron dynamics with QCD as an underlying
force. Moreover the QNJLM
may even be viewed as more natural than the extensions to
more generations, more elementary Higgses, or to SUSY in the SM since the
particles involved in DCSB (with masses of order of the
electroweak scale) belong  in this context only to the ground
state spectrum. In the following sections we shall present
an extension of the SM  with two-Higgs bosons of the QNJLM
where one of the Higgses is a radial excitation of another one.

\section{Higgs Bosons as  Radial Excitations - Model I}

\subsection{Effective potential in Model I}

Let us construct now the  two-flavor
quark models with quasilocal interaction in which the
$ t$- and $ b$-quarks are involved in the DCSB.
In accordance with the SM, the  left components of both  quarks form
a doublet:
\be
  q_{L}=\left( t_{L} \atop b_{L}\right),
\ee
which transforms under
$ SU(2)_{L} $ group as a fundamental representation
while the right components
$ t_{R}, b_{R} $ are singlets.

The Model I which is to satisfy the FCNC suppression has the following
Lagrangian:
\ba
{\cal L_{J}}&=&\bar q_{L}\dirak q_{L} + \bar t_{R}\dirak t_{R}+
  \bar b_{R}\dirak b_{R}+\no
&&
\frac{8\pi^2}{N_{c}\Lambda^{2}}\sum_{k,l=1}^{2}a_{kl}\left(
g_{t, k}J^{T}_{t, k}+g_{b, k}\widetilde J^{T}_{b, k}\right) i\tau_{2}
\left(g_{b, l}J_{b, l}-g_{t, l}\widetilde J_{t, l} \right). \label{lag}
\ea
Here we have introduced the
denotations for doublets of fermion currents:
\be
  J_{t, k}\equiv \bar t_{R}f_{t,k}\left(
-\frac{\partial^{2}}{\Lambda^{2}} \right)q_{L} , \qquad
  J_{b, k}\equiv \bar b_{R}f_{b,k}
\left(-\frac{\partial^{2}}{\Lambda^{2}} \right)q_{L},\label{J}
\ee
and the tilde in
$\widetilde J_{t, k}$  and  $ \widetilde J_{b, k} $ marks
charge conjugated quark currents, roteted with
$ \tau_{2} $ Pauli matrix
\be
\tilde J_{t, k}=i\tau_{2}J^{\star}_{t, k}, \qquad
\widetilde J_{b, k}=i\tau_{2}J^{\star}_{b, k}
\ee
 The subscripts
$ t, b$ indicate right components of
$ t $ and $ b $ quarks in the currents, the index
$ k$ enumerates the formfactors:
\ba
 && f_{t,1}=2-3\left(-\frac{\partial^{2}}{\Lambda^{2}}\right),
\quad f_{t,2}=-\sqrt{3}
\left(-\frac{\partial^{2}}{\Lambda^{2}} \right), \nonumber\\
&& f_{b,1}=2-3\left(-\frac{\partial^{2}}{\Lambda^{2}}\right),
\quad f_{b,2}=-\sqrt{3}\left(-\frac{\partial^{2}}{\Lambda^{2}} \right).\label{ff}
\ea
As the spinor
indices are contracted to each other in  (\ref{J}),
$ J_{t, k} $ transforms as a doublet under
$ SU(2)_{L} $.
$ \tau_{2} $ is a Pauli matrix in the adjoint representation
of the group
$ SU(2)_{L} $.
Coupling constants of the four-fermion interaction are represented by
$ 2\times 2 $ matrix
$ a_{kl} $  and contributed also from
the Yukawa constants $ g_{k, l}, g_{b, k} $.

The Lagrangian density of the Model I (\ref{lag})
to describe the dynamics of composite Higgs bosons
can be obtained by means of
introduction of auxiliary bosonic variables and by integrating out
fermionic degrees of freedom. According to this scheme, we
define two scalar $SU(2)_{L}$-isodoublets:
\be
 \Phi_{1}=\left(\phi_{11}\atop \phi_{12}\right),\qquad
  \Phi_{2}=\left(\phi_{21}\atop \phi_{22}\right)
\ee
and their charge conjugates:
\be
  \widetilde \Phi_{1}=\left(\phi_{12}^{\star}\atop -\phi_{11}^{\star}\right),\qquad
  \widetilde \Phi_{2}=\left(\phi_{22}^{\star}\atop -\phi_{21}^{\star}\right).
\ee
In terms of auxiliary fields, the Lagrangian (\ref{lag}) can be
rewritten in the following way:
\be
{\cal L_{J}} =  L_{kin} +
\frac{N_{c}\Lambda^{2}}{8\pi^2}\sum_{k,l=1}^{2}\Phi^{\dagger}_{k}
(a^{-1})_{kl}\Phi_{l}+i \sum_{k=1}^{2}\left[
g_{t, k}\widetilde \Phi_{k}^{\dagger}J_{t, k}+
g_{b, k}\Phi_{k}^{\dagger}J_{b, k}\right] + h.c.
\ee
The integrating out of  fermionic degrees of freedom will produce
the effective action for Higgs bosons of which we shall keep  only the
kinetic term and  the effective potential
consisting of two- and four-particles
vertices.
The omitted terms are supposedly small,
being proportional to inverse powers of a large scale factor $\Lambda$.
%%%%%%%%%%%%%%%%%%%%%%%%%%%%%%%%%%%%%%%%%%%%%%%%%%%%%%%%%%%%%%%%
The Yukawa constants are chosen of the form
\be
 g_{t,k}=1 ; \qquad g_{b,k}=g
\ee
for $k=(1,2)$. The first constant is set to unit because the
fields
$ \Phi_{1} $ and $ \Phi_{2} $ can always be multiplied by an arbitrary
factor which is absorbed by four fermion coupling constants through
redefinition of polycritical coupling constants. The remaining constant
$ g $ induces thereby the quark mass ratio
$ m_{b}/m_{t} $.
%%%%%%%%%%%%%%%%%%%%%%%%%%%%%%%%%%%%%%%%%%%%%%%%%%%%%%%%%%%%%%%%
The effective potential for the Model I has the following form:
\ba
V_{eff}&=&
\frac{N_{c}}{8\pi^2}\Biggl(
-\sum_{k,l=1}^{2}(\Phi^{\dagger}_{k}\Phi_{l})\Delta_{kl}+
8(\Phi^{\dagger}_{1}\Phi_{1})^{2}\left(
\ln\frac{\Lambda^{2}}{4(\Phi^{\dagger}_{1}\Phi_{1})}+\frac12\right)+
\nonumber\\
&&8g^{4}(\Phi^{\dagger}_{1}\Phi_{1})^{2}\left(
\ln\frac{\Lambda^{2}}{4g^{2}(\Phi^{\dagger}_{1}\Phi_{1})}+\frac12\right)+
\nonumber\\
&&-\frac{159(1+g^{4})}{8}(\Phi^{\dagger}_{1}\Phi_{1})^{2}
+\frac{9(1+g^{4})}{8}(\Phi^{\dagger}_{2}\Phi_{2})^{2}
+\frac{3(1+g^{4})}{4}(\Phi^{\dagger}_{1}\Phi_{1})(\Phi^{\dagger}_{2}\Phi_{2})+
\nonumber\\
&&+\frac{3(1+g^{4})}{4}(\Phi^{\dagger}_{1}\Phi_{2})(\Phi^{\dagger}_{2}\Phi_{1})+
\frac{3(1+g^{4})}{8} (\Phi^{\dagger}_{1}\Phi_{2})^{2}+
\frac{3(1+g^{4})}{8} (\Phi^{\dagger}_{2}\Phi_{1})^{2}-
\nonumber\\
&&-\frac{5\sqrt{3}(1+g^{4})}{4}
(\Phi^{\dagger}_{1}\Phi_{1})\Biggl[(\Phi^{\dagger}_{1}\Phi_{2})+
(\Phi^{\dagger}_{2}\Phi_{1})\Biggr]+\nonumber\\
&&+\frac{\sqrt{3}(1+g^{4})}{4}(\Phi^{\dagger}_{2}\Phi_{2})
\Biggl[(\Phi^{\dagger}_{1}\Phi_{2})+(\Phi^{\dagger}_{2}\Phi_{1})\Biggr]
\Biggr)+O\left(\frac{\ln\Lambda}{\Lambda^{2}}\right),\label{Vmod1}
\end{eqnarray}
where the bilinear ``mass'' term is in general non-diagonal  and  represented
by the real, symmetric
$ 2\times 2 $ matrix
$ \Delta_{kl} $.

We assume the electric charge stability  of vacuum or, in other words,
that only  neutral components  of both Higgs doublets
may have nonzero v.e.v. Hence, one
can deal with only neutral components of the Higgs doublets
in the effective action for studying DCSB.
This part of the Higgs sector
can be investigated separately as a model where two singlets
(not doublets) appear as composite Higgs bosons.
For this purpose, we  use the Quasilocal Two-Channel
model which we have already developed in Sec.3 for
the case of one-flavour\cite{2channel}.

 Following the definitions made in \cite{CP},  we relate the fields
 $ \phi_{1} $,
 $ \phi_{2} $ and $\rho$ to the neutral components of Higgs doublets:

 \begin{equation}
 \phi_{11}=\phi_{1};\qquad \phi_{22}=\phi_{2}+i\rho. \label{parametr}
 \end{equation}
The condition  of minimum of the potential (\ref{Vmod1})
with the charged components of Higgs doublets put to zero values:
$ \phi_{12}=\phi_{21}=0 $, brings the mass-gap equations for them:

 \begin{eqnarray}
\Delta_{11}\phi_{1}+\Delta_{12}\phi_{2}&=&
16\phi_{1}^{3}\ln\frac{\Lambda^{2}}{4\phi_{1}^{2}}+
16g^{4}\phi_{1}^{3}\ln\frac{\Lambda^{2}}{4g^{2}\phi_{1}^{2}}
-(1+g^{4})\Bigl[\frac{159}{4}\phi_{1}^{3}
-\frac{15\sqrt{3}}{4}\phi_{1}^{2}\phi_{2}+
\nonumber\\
&&
+\frac{9}{4}\phi_{1}\phi_{2}^{2}
+\frac{\sqrt{3}}{4}\phi_{2}^{3}
+\frac{\rho^{2}}{4}(3\phi_{1}+\sqrt{3}\phi_{2})\Bigr],\label{gap1}\\
\Delta_{12}\phi_{1}+\Delta_{22}\phi_{2}&=&
(1+g^{4})\Bigl[
-\frac{5\sqrt{3}}{4}\phi_{1}^{3}
+\frac{9}{4}\phi_{1}^{2}\phi_{2}
+\frac{3\sqrt{3}}{4}\phi_{1}\phi_{2}^{2}+
\nonumber\\
&&
+\frac{9}{4}\phi_{2}^{3}
+\frac{\rho^{2}}{4}(\sqrt{3}\phi_{1}+9\phi_{2})\Bigr],\\
4\rho\Delta_{22}&=&\rho(1+g^{4})\left(
3\phi_{1}^{2}+2\sqrt{3}\phi_{1}\phi_{2}+9\phi_{2}^{2}+9\rho^{2}
\right)\label{gap3}
 \end{eqnarray}
Let us consider the equations (\ref{gap1})--(\ref{gap3})
for two cases:
1) $ \rho=0 $
and 2) $ \rho\not=0 $.

When  $ \rho=0 $, assuming that
$ \phi_{1}\not=0 $ and $ \phi_{2}\not=0 $,
we rewrite the equations (\ref{gap1})--(\ref{gap3})
in the following way:
 \begin{eqnarray}
&&\Delta_{11}=16\phi_{1}^{2}\ln\frac{\Lambda^{2}}{4\phi_{1}^{2}}
+16\phi_{1}^{2}g^{4}\ln\frac{\Lambda^{2}}{4g^{2}\phi_{1}^{2}}
+(1+g^{4})\Bigl[-\frac{159}{4}\phi_{1}^{2}
-\frac{15\sqrt{3}}{4}\phi_{1}\phi_{2}+
\nonumber\\
&&\qquad
+\frac94\phi_{2}^{2}+\frac{\sqrt{3}}{4}\frac{\phi_{2}^{3}}{\phi_{1}}\Bigr]-
\Delta_{12}\frac{\phi_{2}}{\phi_{1}},\\
&&\Delta_{22}=(1+g^{4})\Bigl[
\frac{9}{4}\phi_{1}^{2}
+\frac{3\sqrt{3}}{4}\phi_{1}\phi_{2}+
\nonumber\\
&&\qquad
+\frac94\phi_{2}^{2}-\frac{5\sqrt{3}}{4}\frac{\phi_{1}^{3}}{\phi_{2}}\Bigr]-
\Delta_{12}\frac{\phi_{1}}{\phi_{2}}
 \end{eqnarray}
The solution of the mass-gap equation of Gross-Neveu-type is:
\begin{equation}
\phi_1^2\approx
  \frac{ \det\Delta
        }{ 16(1+g^{4})\Delta_{22}
    \ln \left(
          \frac{\Lambda^2}{\mu^2}
        \right)
        } , \quad
\phi_2\approx
  -\frac{ \Delta_{12} }{ \Delta_{22} }\phi_1.
\end{equation}
The solution of the mass-gap equation of the Abnormal-type is:
\begin{equation}
\phi_2^2 \approx \frac4{9(1+g^{4})}\Delta_{22},\quad
\phi_1^2 \approx
        \frac{ \Delta_{22}^{1/3}
                (3\sqrt{3}\Delta_{12}-\Delta_{22})^{2/3}
                }{ 4\cdot 3^{5/3}(1+g^{4})
                \ln^{2/3}
                  \left(
                    \frac{\Lambda^2}{\mu^2}
                  \right)},
\end{equation}
%i..e. easy to see that in this case the solutions, in general, the same ones as
%in the Two-Channel model .

For the case 2, of a non-zero $\rho$ , the mass-gap equations reads:
 \begin{eqnarray}
\Delta_{11}&=&16\phi_{1}^{2}\ln\frac{\Lambda^{2}}{4\phi_{1}^{2}}+
16g^{4}\phi_{1}^{2}\ln\frac{\Lambda^{2}}{4g^{2}\phi_{1}^{2}}\nonumber\\
&&+(1+g^{4})\Bigl[-
\frac{159}{4}\phi_{1}^{2}-\frac{5\sqrt{3}}{2}\phi_{1}\phi_{2}
+\frac{3}{4}\left(\phi_{2}^{2}+
\rho^{2}\right)\Bigr], \label{g1}\\
\Delta_{12}&=&(1+g^{4})\Bigl[
-\frac{5\sqrt{3}}{4}\phi_{1}^{2}+\frac32\phi_{1}\phi_{2}+
\frac{\sqrt{3}}{4}\left(\phi_{2}^{2}+\rho^{2}\right)\Bigr],\\
\Delta_{22}&=&(1+g^{4})\Bigl[
\frac34\phi_{1}^{2}+\frac{\sqrt{3}}{2}\phi_{1}\phi_{2}+
\frac94\left(\phi_{2}^{2}+
\rho^{2}\right)\Bigr] \label{g3}
 \end{eqnarray}
The mass-gap equations (\ref{g1})--(\ref{g3}) can be rewritten
in an equivalent form:
 \begin{eqnarray}
3\Delta_{11}-3\Delta_{22}+9\sqrt{3}\Delta_{12}&=&
48\phi_{1}^{2}\left(\ln\frac{\Lambda^{2}}{4\phi_{1}^{2}}-3\right)+
48g^{4}\phi_{1}^{2}\left(\ln\frac{\Lambda^{2}}{4g^{2}\phi_{1}^{2}}-3\right),\label{gap4}\\
\Delta_{22}-3\sqrt{3}\Delta_{12}&=&(1+g^{4})(12\phi_{1}^{2}-4\sqrt{3}\phi_{1}\phi_{2}),\\
\Delta_{22}&=&(1+g^{4})\left[\frac34\phi_{1}^{2}+\frac{\sqrt{3}}{2}\phi_{1}\phi_{2}+\frac94
\left(\phi_{2}^{2}+\rho^{2}\right)\right].  \label{gap5}
 \end{eqnarray}
From (\ref{gap4})--(\ref{gap5}) it is clear that for fixed
$ \Delta_{kl} $ while
$ \Lambda $ grows larger, the solution exists if $ \Delta_{kl} $ parameters
are chosen close to a particular plane in the parametric space. This
plane is defined by the equation:
 \begin{equation}
\Delta_{22}=3\sqrt{3}\Delta_{12}. \label{hyp}
 \end{equation}
When  $ \Delta_{kl} $ satisfy the equation (\ref{hyp}) exactly,
the solution is found to be as follows (in the large-log approximation):
 \begin{equation}
\phi_{1}^{2}=\frac{\Delta_{11}}{16(1+g^{4})\ln\frac{\Lambda^{2}}{\mu^{2}}}
\left[1+O\left(\frac{1}{\ln\frac{\Lambda^{2}}{\mu^{2}}}\right)\right],
 \end{equation}
 \begin{equation}
\phi_{2}=\sqrt{3}\phi_{1},
 \end{equation}
\begin{equation}
\rho^{2}=\frac49\frac{\Delta_{22}}{(1+g^{4})}
\left[1+O\left(\frac{1}{\ln\frac{\Lambda^{2}}{\mu^{2}}}\right)\right].
\end{equation}
\subsection{Mass spectrum in Model I}

\hspace*{3ex}The mass spectrum of related bosonic states is
determined by the Eqs. (26)-(31) and taking into account the conditions
necessary for a minimum of the potential (52,53). The solutions at
$-m^2 = p^2 < 0$ one can obtain from the Eqs:
\be
\det(Ap^{2}+B)=0, \label{spectrumeqn}
\ee
The "kinetic" matrix $\hat{A}$ as being proportional to $p^{2}$ is derived in the
soft-momentum expansion in powers of $p^{2}$ and  in a large-log$\Lambda$
approximation. Because the expressions for $\hat{A}$ and $\hat{B}$ are
cumbersome we give their explicit form in the Appendix A and B correspondingly.
After substituting expressions for the matrix $\hat{A}, \hat{B}$
into (\ref{spectrumeqn})
one  can get the mass spectrum for the neutral Higgs bosons in Model I.
For the case, $\rho = 0$, the mass spectrum is resembling ones
in Two-Channel model,
in particular, the Gross-Neveu-type solution brings the spectrum for scalars:
\begin{eqnarray}
m_{\sigma'}^2
  & \approx &
        -\frac{ 4\Delta_{22} }{ 3(1+g^{2}) }, \\
m_{\sigma}^2
  & \approx &
        -\frac{
           \det\Delta  }{
           (1+g^{2})\Delta_{22}\ln
                \left(
                   \frac{\Lambda^2}{\mu^2}
                \right) }
%             }
        =4m_{d}^2
\end{eqnarray}
and for pseudoscalars:
\begin{eqnarray}
m_{\pi'}^2
   & \approx & -\frac{ 4\Delta_{22} }{ 3(1+g^{2}) }, \\
m_{\pi}^2
  & = & 0.
\end{eqnarray}

The Abnormal solution induces the following mass spectrum for scalars:
\begin{eqnarray}
m_{\sigma'}^2
  & \approx &
     \frac{ 8\Delta_{22} }{ 3(1+g^{2}) },\\
m_{\sigma}^2
  & \approx &
        \frac{2 \Delta_{22}^{1/3}
          (3\sqrt{3}\Delta_{12}-\Delta_{22})^{2/3} }{
           3^{2/3}(1+g^{2})\ln^{2/3}
                \left(
                 \frac{\Lambda^2}{\mu^2}
                \right)
              }=6m_{d}^2
\end{eqnarray}
and for pseudoscalars:
\begin{eqnarray}
m_{\pi'}^2
   & \approx &
     \frac{ 3^{2/3} (3\sqrt{3}\Delta_{12}-\Delta_{22})^{4/3} }{
        27(1+g^{2}) \Delta_{22}^{1/3} \ln^{1/3}
          \left(
            \frac{\Lambda^2}{\mu^2}
           \right)
          },\\
m_{\pi}^2&= &0
\end{eqnarray}
Remark that dynamical mass $m_{d}$
is in fact the mass of t quark in the Model-I,
because the v.e.v. of $\phi_{11}$, which is parametrized as
$\phi_1 \equiv\langle\phi_{11}\rangle$, gives the value of mass of t-quark.

The mass spectrum in the $P$-parity Breaking Phase,
for a non-zero $\rho$  is:
\begin{eqnarray}
m_{1}^{2}&=&0,\\
m_{2}^{2}&\approx&\frac{3\Delta_{11}-\Delta_{22}}{3(1+g^{2})\ln\frac{\Lambda^{2}}{\mu^{2}}}
        \approx 16\phi_{1}^{2}= 4m_{d}^{2},\\
m_{3}^{2}&\approx &\frac{8\Delta_{22}}{3(1+g^{2})},\\
m_{4}^{2}&\approx &\frac{3\Delta_{11}+7\Delta_{22}}{27(1+g^{2})\ln\frac{\Lambda^{2}}{\mu^{2}}},
\end{eqnarray}

Thus, we have constructed the Model I where:
 \begin{itemize}
\item[a)] Two composite Higgs doublets are created dynamically as a consequence
of DCSB in two channels.
\item[b)]In 2HQ Model I Higgs bosons are rather radial, ground and
excited states in the scalar-pseudoscalar channels.
\item[c)]The appropriate fine tuning leads also to spontaneous breaking of
P-parity and, therefore, of CP-parity in the Higgs sector.
 \end{itemize}

\section{Top-Bottom Condensation for 2HQM Model- II}

\subsection{Effective potential in Model II}

\hspace{5mm}The Lagrangian density of the Model II to describe the
dynamic of two composite Higgs bosons which consist of bound states
(condensates $\bar tt,\bar bb$) and satisfy the FCNC suppression\cite{GW} can be written  as:
\be
{\cal L_{J}} =  L_{kin} +
\frac{N_{c}\Lambda^{2}}{8\pi^2}\sum_{k,l=1}^{2}\Phi^{\dagger}_{k}
(a^{-1})_{kl}\Phi_{l}+i\bar q(\widetilde {\mbox{\bf M}}P_L+
        \widetilde {\mbox{\bf M}}^{\dagger}P_R )q  + h.c.,
\ee
where $P_{L(R)}=1/2(1\pm\gamma_5)$ - the left and right projectors,
and $\leftvertex$ is the two-by-two flavour matrix:

\be
\leftvertex=
\sum_{m=1}^2
\left(\begin{array}{cc}
\phimtwo\,\displaystyle\fitm\stick & -\phimone\,\displaystyle\fitm \\
\phimone^{\star}\,\displaystyle\fibm\stick & \phimtwo^{\star}\,
\displaystyle\fibm
\end{array}\right),                          \label{mmatrix}
\ee
where we set for the Yukawa coupling constants $g_{t,k}, g_{b,k} = 1$
(two Yukawa constants are due to  the renormalization of Higgs fields and other ones we
choose equal one). In this Model II $\Phi_{1}, \Phi_{2}$
give masses to $ up-,down-$type quarks.
The structure of quark interaction is specified in four formfactors:
\ba
f_{t,1}&=&1-\ctone\frac{\partial^{2}}{\Lambda^{2}}, \nonumber \\
f_{t,2}&=&-\cttwo\frac{\partial^{2}}{\Lambda^{2}}, \nonumber \\
f_{b,1}&=&-\cbone\frac{\partial^{2}}{\Lambda^{2}}, \nonumber \\
f_{b,2}&=&1-\cbtwo\frac{\partial^{2}}{\Lambda^{2}}.
        \label{formfactors}
\ea

When the  chiral symmetry is broken, the v.e.v. of neutral
Higgs fields are non-zero and
the true Yukawa vertices should be obtained by subtracting from
$\leftvertex$ its v.e.v.
\be
        \mbox{\bf\cal M}=\leftvertex-\mavr , \label{mmatrix2}
\ee
where
$\langle\dots\rangle $ means a v.e.v.:
\be
        \langle\leftvertex\rangle=\left(
        \begin{array}{cc}
            m_{t} & 0 \\
            0 & \faza{}m_{b}
          \end{array}
        \right).\label{mmatrix3}
\ee
The elements of the matrix (\ref{mmatrix3}) are the quark mass functions:
\ba
&&m_{t}(\tau)=\phi_{1}(1+c_{t,1}\tau) + \phi_{2}\faza{}c_{t,2}\tau,\nonumber\\
&&m_{b}(\tau)=\phi_{1}c_{b,1}\tau + \phi_{2}\faza{}(1+c_{b,2}\tau),\qquad
\tau \equiv - \frac{\partial^2}{\Lambda^2} ,
\ea
defined to be real and $\phi_{1} = <\phi_{12}>, \phi_{2} = <\phi_{22}>$.
The non-zero phase at
$ m_{b} $, which is displayed explicitly
in (\ref{mmatrix3}),
may appear if the v.e.v. of
$ \phi_{22} $ acquires irremovable phase factor when the chiral symmetry is
broken.

As the vacuum charge stability is assumed, $\mavr$ is diagonal,
so $\mavr$ and $\mavr^{\dagger}$ commute and can be placed in any order in products
of themselves.

The effective potential of composite two-Higgs model II in which
the interaction of quarks and Higgs bosons is described by
formfactors (\ref{formfactors}) reads:

\ba
V_{eff}&=&-\Delta_{11}\phi_{1}^{2}-
        2\Delta_{12}\phi_{1}\phi_{2}\cos\delta_{0}-
        \Delta_{22}\phi_{2}^{2}+
        \frac12\phi_{1}^{4}\left(\ln\frac{\Lambda^{2}}{\phi_{1}^{2}}+
        \frac12\right)+\nonumber\\
     &&\quad+\frac12\phi_{2}^{4}\left(\ln\frac{\Lambda^{2}}{\phi_{2}^{2}}+
        \frac12\right)+
        \frac12\bigl(
        J_{1111}\phi_{1}^{4}+4J_{1112}\phi_{1}^{3}\phi_{2}\cos\delta_{0}+
        \nonumber\\
        &&\quad+2(J_{1122}+J_{1221}+
                J_{1212}\cos2\delta_{0})\phi_{1}^{2}\phi_{2}^{2}+
        4J_{1222}\phi_{1}\phi_{2}^{3}\cos\delta_{0} +\nonumber\\
        &&\quad+J_{2222}\phi_{2}^{4}
        \bigr)+O\left(\frac{1}{\Lambda^{2}}\right)
        \label{Veff},
\ea
where $v.e.v.$ of fields are: $<\phi_{12}> = \phi_{1}, <\phi_{22}> =\phi_{2}$.
(For the concrete choice of formfactors in Model II the form of the effective
potential is displayed in the Appendix C)
Its minimum is described by solutions of the mass-gap equations (21), (22)
which for the Model II are:
\ba
        &&2\Delta_{11}\phi_{1}+
        2\Delta_{12}\phi_{2}\cos\delta_{0}=\lefteqn\nonumber\\
        &&\qquad=2\phi_{1}^{3}\ln\left(\frac{\Lambda^{2}}{\phi_{1}^{2}}
                \right)+2J_{1111}\phi_{1}^{3}+6J_{1112}\phi_{1}^{2}\phi_{2}
                \cos\delta_{0}+2J_{1222}\phi_{2}^{3}\cos\delta_{0}+\nonumber\\
                &&\qquad
                +2\left(J_{1122}+J_{1221}+
                J_{1212}\cos2\delta_{0}\right)\phi_{1}\phi_{2}^{2},
                \label{massgapfirst}\\
        &&2\Delta_{12}\phi_{1}\phi_{2}\sin\delta_{0}=\lefteqn\nonumber \\
        &&\qquad= 2J_{1112}\phi_{1}^{3}\phi_{2}\sin\delta_{0}+
                2J_{1212}\phi_{1}^{2}\phi_{2}^{2}\sin2\delta_{0}+
                J_{1222}\phi_{1}\phi_{2}^{3}\sin\delta_{0},\\
        &&2\Delta_{22}\phi_{2}+
        2\Delta_{12}\phi_{1}\cos\delta_{0}=\lefteqn\nonumber\\
        &&\qquad=2\phi_{2}^{3}\ln\left(\frac{\Lambda^{2}}{\phi_{2}^{2}}
                \right)+2J_{2222}\phi_{2}^{3}+6J_{1222}\phi_{1}\phi_{2}^{2}
                \cos\delta_{0}+2J_{1112}\phi_{1}^{3}\cos\delta_{0}+\nonumber\\
                &&\qquad+
                2\left(J_{1122}+J_{1221}+
                J_{1212}\cos2\delta_{0}\right)\phi_{1}^{2}\phi_{2},
                \label{massgaplast}
\ea
where
$ J_{klmn}\; (k,l,m,n=1,2) $ are the integrals:
\ba
J_{klmn}&=&
 \int\limits_{0}^{1}\left(
f_{t,k}(\tau)
f_{t,l}(\tau)
f_{t,m}(\tau)
f_{t,n}(\tau)+ \right. \nonumber\\
&&\left. +
f_{b,k}(\tau)
f_{b,l}(\tau)
f_{b,m}(\tau)
f_{b,n}(\tau)+\right.\nonumber\\
&&\left.+
f_{t,k}(\tau)
f_{b,l}(\tau)
f_{b,m}(\tau)
f_{t,n}(\tau)-\right.\nonumber \\
&&\left.-
f_{t,k}(\tau)
f_{t,l}(\tau)
f_{b,m}(\tau)
f_{b,n}(\tau)+\right.\nonumber\\
&&\left.+
f_{b,k}(\tau)
f_{b,l }(\tau)
f_{t,m}(\tau)
f_{t,n}(\tau)-\right.\nonumber\\
&&\left.-
f_{b,k}(\tau)
f_{t,l}(\tau)
f_{t,m}(\tau)
f_{b,n}(\tau)-\right.\nonumber\\
&&\left.-
f_{t,k}(0)
f_{t,l}(0)
f_{t,m}(0)
f_{t,n}(0)-\right.\nonumber\\
&&\left.-
f_{b,k}(0)
f_{b,l}(0)
f_{b,m}(0)
f_{b,n}(0)-\right.\nonumber\\
&&\left.-
f_{t,k}(0)
f_{b,l}(0)
f_{b,m}(0)
f_{t,n}(0)+\right.\nonumber\\
&&\left.+
f_{t,k}(0)
f_{t,l}(0)
f_{b,m}(0)
f_{b,n}(0)-\right.\nonumber\\
&&\left.-
f_{b,k}(0)
f_{b,l}(0)
f_{t,m}(0)
f_{t,n}(0)+\right.\nonumber\\
&&\left.+
f_{b,k}(0)
f_{t.l}(0)
f_{t,m}(0)
f_{b,n}(0)
\right)\frac{d\!\tau}{\tau}.
\ea

It is more convenient to
solve the equations (\ref{massgapfirst})--(\ref{massgaplast})
for the variables
$ \Delta_{lm} $ rather than
$ \phi_{1},\;\phi_{2},\; \delta_{0} $. The variables
$ \phi_{1},\;\phi_{2},\; \delta_{0} $ will be treated as
input parameters while $ \Delta_{lm} $ as the unknowns. The reason for
this is that we do not know $ \Delta_{lm} $ from any global theory,
we rather fit them so that $ \phi_{1},\;\phi_{2},\; \delta_{0} $
conform to experiment. The equations
(\ref{massgapfirst})--(\ref{massgaplast}) are linear
for $ \Delta_{lm} $ and can easily be solved; one just expresses
$ \Delta_{lm} $ through $ \phi_{1},\;\phi_{2},\; \delta_{0} $ and
substitutes them in every place they appear.
As usual two cases must be considered separately.

\noindent 1)For
$ \delta_{0}=0 $:
\ba
\Delta_{11}&=&\phi_{1}^{2}\ln\left(\frac{\Lambda^{2}}{\phi_{1}^{2}}\right)
        +J_{1111}\phi_{1}^{2}+3J_{1112}\phi_{1}\phi_{2}+\nonumber \\
        &&+
        \left(
        J_{1122}+J_{1221}+J_{1212}
        \right)\phi_{2}^{2}+J_{1222}\frac{\phi_{2}^{3}}{\phi_{1}}-
        \Delta_{12}\frac{\phi_{2}}{\phi_{1}}\label{delta0begin},\\
\Delta_{22}&=&\phi_{2}^{2}\ln\left(\frac{\Lambda^{2}}{\phi_{2}^{2}}\right)
        +J_{2222}\phi_{2}^{2}+3J_{1222}\phi_{1}\phi_{2}+\nonumber \\
        &&+
        \left(
        J_{1122}+J_{1221}+J_{1212}
        \right)\phi_{1}^{2}+
        J_{1112}\frac{\phi_{1}^{3}}{\phi_{2}}-
        \Delta_{12}\frac{\phi_{1}}{\phi_{2}},\label{delta0end}
\ea
where
$ \Delta_{12},\; \phi_{1},\; \phi_{2} $ are treated as input parameters;

\noindent  2) and for
$ \delta_{0}\not=0 $:
\ba
\Delta_{11}&=&
        \phi_{1}^{2}\ln\left(\frac{\Lambda^{2}}{\phi_{1}^{2}}\right)+
        J_{1111}\phi_{1}^{2}+2J_{1112}\phi_{1}\phi_{2}\cos\delta_{0}+
        \nonumber \\
        &&+
        \left(J_{1122}+J_{1221}-J_{1212}\right)\phi_{2}^{2},
                \label{deltabegin}\\
\Delta_{22}&=&
        \phi_{2}^{2}\ln\left(\frac{\Lambda^{2}}{\phi_{2}^{2}}\right)+
        J_{2222}\phi_{2}^{2}+2J_{1222}\phi_{1}\phi_{2}\cos\delta_{0}+
        \nonumber \\
        &&+
        \left(J_{1122}+J_{1221}-J_{1212}\right)\phi_{1}^{2},\\
\Delta_{12}&=&
        J_{1112}\phi_{1}^{2}+J_{1222}\phi_{2}^{2}+
                2J_{1212}\phi_{1}\phi_{2}\cos\delta_{0}. \label{deltaend}
\ea
The mass spectrum of related bosonic states is determined by the matrices
$\hat{A } $ and $\hat{B}$
of the second variations of the effective potential ( \ref{Veff}) (see
Appendix A, B).

\subsection{Mass spectrum in Model II}

\hspace*{3ex}After substituting explicit forms for the $\hat{A,}\, \hat{B}$ into
(\ref{spectrumeqn}), one can
obtain the mass-spectrum for the composite neutral Higgses in Model II.

\noindent 1) For
$ \delta_{0}=0 $:
\ba
m_{\sigma'}&\approx& 2m_{t},\nonumber\\
m_{\pi}&=&0,\nonumber\\
m_{\sigma}&\approx&\sqrt{2(\frac{\Delta_{12}-\mt{2}J_{1112})}{r}},\nonumber\\
m_{\pi'}&\approx&m_{\sigma'},
\ea
where
\be
r=\frac{\mb{}}{\mt{}}\lgt\sim 1.
\ee
If the ratio were (say, for the fourth generation):
\be
\frac{\mb{}}{\mt{}}= 0(1)\,\quad (\Lambda\rightarrow\infty),
\ee
than one would get:
\ba
m_{\sigma}&=&2m_{b},\nonumber\\
m_{\pi'}^2&=&\frac{2(\mt{2}+\mb{2})}{\mt{}\mb{}\lgt}\times\nonumber\\
&&\times \biggl(\Delta_{12}-\mt{2}J_{1112}-2\mt{}\mb{}J_{1212}-\mb{2}J_{1222}\biggr).
\ea

\noindent 2) For
$ \delta_{0}\not=0 $:
\ba
m_{1}&=&0,\\
m_{2}&\approx& 2\mt{},\\
m_{3}&\approx&
\mt{}\sqrt{\frac{J_{1212}}{\lgt}}\\
m_{4}&\approx&
2r\mb{}|\sin\delta_{0}|\sqrt{J_{1212}}\, .
\ea
For the case of
$ \delta_{0}\not=0 $, the model predicts a low  value
for $ m_{4} $.

If one considered (that may take place for the fourth generation):
\be
 \mt{}\sim \mb{}\sim1\quad (\Lambda\rightarrow\infty),
\ee
 the mass-spectrum would turn out to be as follows:
\ba
m_{1}&=&0\\
m_{2}&\approx& 2\mt{}\\
m_{3}&\approx& 2\mb{}\\
m_{4}&\approx& 2|\sin\delta_{0}|\sqrt{\frac{J_{1212}(\mt{2}+\mb{2})}{\lgt}}\, .
\ea
We notice that when $ \delta_{0}\not= 0 $ we have neither scalars no pseudoscalars
 any longer
because the particles which are eigenstates of the energy operator,
are mixed of both P-even parity and P-odd parity fields, hence the
former classification
by parity does not hold for this particular case.

\subsection{Mass spectrum in Model II for the special choice
of Yukawa couplings with FCNC suppression}

The suppression of the  flavor changing neutral currents
in a composite two-Higgs model can be
implemented  when Yukawa couplings are chosen
as follows:
\be
g_{t,1}=g_{b,2};\qquad g_{t,2}=g_{b,1}=\sqrt{5}.
\ee
It can be easily seen from the potential in Appendix C that
with the above choice of
$ g_{t,k}, g_{b,k} $
the last two terms in (191) come to zero. This prevents the
appearence of the Yukawa coupling inducing FCNC
 from radiative corrections.
In this case the potential of Model type II reads:

\begin{eqnarray}
V_{eff}&=&
{{N_{c}}\over{8\pi^2}}\Biggl\{
  -\Delta_{11}(\Phi_{1}^{\dagger}\Phi_{1})
  -\Delta_{12}(\Phi_{1}^{\dagger}\Phi_{2})
  -\Delta_{21}(\Phi_{2}^{\dagger}\Phi_{1})
  -\Delta_{22}(\Phi_{2}^{\dagger}\Phi_{2})+\nonumber\\
&&+\frac{\nu_{+}^{2}(0)}{2}\left(\ln{{\Lambda^{2}}\over{\nu_{+}(0)}}+\frac12\right)+
\frac{\nu_{-}^{2}(0)}{2}\left(\ln{{\Lambda^{2}}\over{\nu_{-}(0)}}+\frac12\right)+
\nonumber\\
&&+\frac{33}{4}(\Phi_{1}^{\dagger}\Phi_{1})^{2}+
\frac{33}{4}(\Phi_{2}^{\dagger}\Phi_{2})^{2}+
\nonumber\\
&&+\frac{33}{2}(\Phi_{1}^{\dagger}\Phi_{1})(\Phi_{2}^{\dagger}\Phi_{2})
-\frac32
(\Phi_{1}^{\dagger}\Phi_{2})(\Phi_{2}^{\dagger}\Phi_{1})+
\nonumber\\
&&+\frac{15}{4}\left[(\Phi_{1}^{\dagger}\Phi_{2})^{2}+
(\Phi_{1}^{\dagger}\Phi_{2})^{2}\right]
\Biggr\}
\end{eqnarray}
where for
$ \nu_{\pm}(0) $ we have
\begin{eqnarray}
&&\nu_{\pm}=
2(\Phi_{1}^{\dagger}\Phi_{1})+
2(\Phi_{2}^{\dagger}\Phi_{2})\pm
2\Bigl[
(\Phi_{1}^{\dagger}\Phi_{1})^{2}+
(\Phi_{2}^{\dagger}\Phi_{2})^{2}+
\nonumber\\
&&\qquad\pm
2(\Phi_{1}^{\dagger}\Phi_{1})
(\Phi_{2}^{\dagger}\Phi_{2})-
4(\Phi_{1}^{\dagger}\Phi_{2})
(\Phi_{2}^{\dagger}\Phi_{1})
\Bigr]^{1/2}
\end{eqnarray}

To be short we skip the calculations and
display only the mass spectrum.
When the mass ratio of
$ t$- and
$ b$-quarks $ m_{t}/m_{b} $ holds unchanged while
$ \Lambda $ goes to infinity,  in the large-log approach
we get
the following estimations for Higgs' masses:

\begin{enumerate}
\item for the case of concerved P-parity ($ \delta_{0}=0 $):
\ba
m_{\sigma}&=&2m_{b},\nonumber\\
m_{\sigma'}&\approx& 2m_{t},\nonumber\\
m_{\pi}&=&0,\nonumber\\
m_{\pi'}&=&\sqrt{\frac{(\mt{2}+\mb{2})}{2\mt{}\mb{}\lgt}
\biggl(\Delta_{12}-\frac{15}{4}\mt{}\mb{}\biggr)}.
\ea

\item and for the P-parity breaking phase:
\ba
m_{1}&=&0\\
m_{2}&\approx& 2\mt{}\\
m_{3}&\approx& 2\mb{}\\
m_{4}&\approx& |\sin\delta_{0}|\sqrt{\frac{15(\mt{2}+\mb{2})}{8\lgt}}\, .
\ea
\end{enumerate}

\bigskip

Thus in the Model II the Quasilocal Yukawa interaction with Higgs
doublets reduces at low energies to a conventional local one where
each Higgs doublet couples to a definite charge current and its $v.e.v.$
brings the mass either to $up$- or to $down$- components of fermion
doublets.
Based on the FCNC suppression, the Model II leads to the relation
$m_{t}>>m_{b}$ and so to an enhanced coupling of the light scalar
(pseudoscalar) boson to the $down$-type quarks while suppressing
the coupling to the $up$-type quarks.
The Model II has a  broad spectrum of excited bound states which is to
parametrize the data, in particular, obtained from the Next Linear Collider.

\section{Summary}
In our paper we have proposed a set of Quasilocal NJL-type quark
models (QNJLM) which lead to a larger spectrum of ground and
excited states in the polycritical regime. From the viewpoint of the
SM, these models are considered as more natural than common
extensions of the SM, since they do not enlarge the number of
elementary particles in fermionic sector and  preserve the
symmetries of the SM. For the toy
Two-Channel Quasilocal quark model, near tricritical point we have
found three major phases: a symmetrical one and two phases
with DCSB, different in correlation lengths in scalar channels.
On a particular plane in the space of coupling constants we
discovered the special $P$-parity breaking phase.
It means that in
such a phase there exist heavy scalar states which can
decay into two or three light pseudoscalars. This phenomenon
of dynamical $P$-parity breaking can be used in the extensions
of the SM where several Higgs bosons are composite ones.
In the framework of the QNJLM we have presented two
Models which provide at low energies two composite Higgs
doublets, as minimal extensions of the Top-Mode Standard
Model \cite{Bardeen},\cite{Lind}. In the 2HQ Model I Higgs bosons
are rather radial, ground and excited states in the scalar
-pseudoscalar channels. In the 2HQ Model II, which
consistent with the requirement of natural flavour conservation
\cite{GW}, strong forces lead to the formation of  top and
bottom bound states (and corresponding condensates) and generate
masses of $t$,$b$-quarks. In Model II we have concentrated
on the scenario where each of the neutral components of the
two doublets $\phi_{1,2}$ (with v.e.v. $v_{1,2}$) couple
at low energies respectively to
the $I_{3}=\pm\frac{1}{2}$ fermion fields
The FCNC suppression leads to the relation $m_{t}>>m_{b}$
and to an
enhanced coupling of the light scalar (pseudoscalar) boson to
the $down$-type quarks and the charged leptons while
suppressing the coupling to the $up$-type quarks.
The existence of light neutral Higgs (pseudo)scalar
bosons in the framework of 2HDM is not excluded by
existing data ($< 40\, GeV$). The chance that it can
be seen at the Next Linear Collider in the $\gamma\gamma$
processes has been pointed out in \cite{Skjold},\cite{ChKr}.
As a result of complexity of
two v.e.v.'s for two composite Higgs doublets the
dynamical $CP$-violation may appear in the Higgs sector. At high
energies these channels are strongly coupled and
one could say that two-composite Higgs doublets
partially represent the mixture with excited states.
If such excited states exist then they will modify
the Higgs mass predictions. In addition, we remark
that low values for the Higgs masses of the additional
states could actually change the window
for $M_{H}$ since these states could give a
significant contribution to the $\rho$-parameter
\cite{Lang}. From our consideration
we have seen that the appearance of dynamical
$CP$-violation in the Higgs sector
imposes strong bounds on Higgs masses, in particular,
one light scalar Higgs boson is unavoidable.
The experimental implications of such effects are expected
to be rather small in the fermion sector of the SM \cite{Gunion},\cite{Skjold}.
These effects are observable in decays of heavy
Higgs particles (namely, pseudoscalar Higgses may
decay into scalar ones, scalar Higgs may decay
into pseudoscalar ones) and in decays of Higgses
particles into two vector bosons where $CP$-even and
$CP$-odd amplitudes appear. At high energies
the appearance of the appreciable $CP$-violation
could be important both as a source of electron
and neutron electric dipole moments \cite{Barr}
and as a mechanism for EW scale baryogenesis\cite{Larry},
\cite{Cline}.
Besides one expect also that modifications
of the SM Lagrangian (the Higgs and Top interactions)
by higher dimensional vertices may enhance the
Higgs production at hadron colliders \cite{Diaz}.

The theory of two composite Higgs bosons
which we have discussed in our paper should be regarded
as a viable alternative to other approaches to the BSM
and perhaps the major progress in the alternative
approaches will come when the first direct experimental
results associated with the origin of EWSB begin to
appear.

The purpose of this paper has been to
elaborate the very design of  quasilocal NJL-quark models
with two-composite Higgs bosons. A more comprehensive
analysis of low-energy particle characteristics in these models
is postponed to the next paper in this series of.The numerical computation
of bounds on mass spectra, Yukawa coupling constants and
decay widths  with taking into account the renormalization-group
corrections will be presented elsewhere.

\section{Acknowledgements}
We express our gratitude to DFG for financial support which make it
 possible to prepare this paper and also for giving  us the opportunity to
 continue  our research on
properties of composite Higgs bosons. We also thank very much
G. Kreyerhoff for stimulating
discussions and  for his help and encouragement.

This work has been
supported by  RFBR (Grant No. 95-02-05346a),
by  INTAS (Grant No. 93-283ext), GRACENAS
(Grant No. 95-6.3-13) and
 by DFG (Grant No. 436 RUS 113/227/1 (R).
\vfill\newpage

\section*{Appendix A: Kinetic matrix $\hat{A}$ for composite two-Higgs bosons}

\hspace*{3ex} In this appendix we calculate the kinetic term for
composite Two-Higgs  Quasilocal Quark Models which is
obtained by calculation of the one-loop diagram:

\begin{figure}[h]
\unitlength=1mm
\special{em:linewidth 0.4pt}
\linethickness{0.4pt}
\begin{picture}(91.00,30.00)
\put(72.00,16.00){\circle{14.00}}
\put(79.00,16.00){\circle*{2.00}}
\put(65.00,16.00){\circle*{2.00}}
\put(79.00,16.50){\line(1,0){12.00}}
\put(79.00,15.50){\line(1,0){12.00}}
\put(53.00,16.50){\line(1,0){12.00}}
\put(53.00,15.50){\line(1,0){12.00}}
\put(72.00,28.00){\makebox(0,0)[cc]{$k+{p \over 2}$}}
\put(72.00,3.00){\makebox(0,0)[cc]{$k-{p\over 2}$}}
\put(55.00,20.00){\makebox(0,0)[cc]{$\overrightarrow p$}}
\put(90.00,20.00){\makebox(0,0)[cc]{$\overrightarrow p$}}
\put(72.00,23.00){\makebox(0,0)[cc]{$>$}}
\put(72.00,9.00){\makebox(0,0)[cc]{$<$}}
\end{picture}
\caption{One-loop diagram for calculation of kinetic term.}
\label{diagram}
\end{figure}
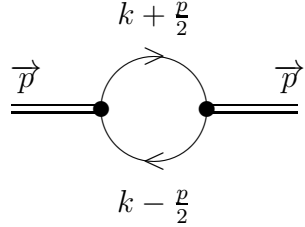\noindent
Here $p$ is an  incoming momentum, and
$k$ is a momentum running around the loop.

The loop diagram (Fig.1)  gives the following expression:
\ba
       && -\frac12\int
        \limits_{|k|<\Lambda}\!\frac{d^4\!k}{(2\pi)^4}
        \mbox{\rm Tr}\Biggl[
                \left(
                i\mbox{\bf\cal M}P_L+i\mbox{\bf\cal M}^{\dagger}P_R\right)
                \mbox{\bf\cal D}^{-1}\left(\left(k+
                        \frac{p}{2}\right)^2\right)\nonumber \\
         &&\quad\times\left(i\mbox{\bf\cal M}P_L+i\mbox{\bf\cal M}^{\dagger}P_R\right)
                \mbox{\bf\cal D}^{-1}\left(\left(k-\frac{p}{2}\right)^2\right)
        \Biggr].\label{integral}
\ea
Here the coefficient $1/2$ is due to  symmetry of the diagram
(Fig.1).
The full expression within the square brackets in (\ref{integral})
is an element of a direct
product of three spaces: color, flavor and spinor and the trace operation
is related to all them.
We define the vertex:
\unitlength=1mm
\special{em:linewidth 0.4pt}
\linethickness{0.4pt}
\begin{picture}(28.00,11.00)(0,10)
\put(4.00,3.00){\line(1,0){21.00}}
\put(14.00,3.00){\circle*{2.00}}
\put(13.50,3.00){\line(0,1){8.00}}
\put(14.50,3.00){\line(0,1){8.00}}
\put(16.50,2.00){$ < $}
\put(8.50,2.00){$ < $}
\put(50.00,7.00){\makebox(0,0)[cc]{
       $=i(\mbox{\bf M}P_L+\mbox{\bf M}^{\dagger}P_R$)

}}
\end{picture}
\vspace{15mm}

\noindent and the fermion propagator:
\vskip 10mm

\unitlength=1mm
\special{em:linewidth 0.4pt}
\linethickness{0.4pt}
\begin{picture}(29.00,7.00)
\put(7.00,7.00){\line(1,0){18.00}}
\put(16.00,7.00){\makebox(0,0)[cc]{$<$}}
\put(29.00,9.00){\makebox(0,0)[lt]{\vbox{
\hbox{$=\mbox{\bf\cal D}^{-1}(q)=(\hat q+i \mavr P_L+i\mavr^{\dagger}P_R)^{-1}=$}
        \hbox{$(\hat q-i\mavr P_L-i\mavr^{\dagger}P_R)\Delta(q^2),$}}}}
\end{picture}
\vskip 10 mm

\noindent where  $\Delta$ is a matrix function:
\be
        \Delta(q^2)=(q^2+\mavr \mavr^{\dagger})^{-1}=
   \left(
    \begin{array}{cc}
           \frac{\displaystyle 1\strut}{
        \displaystyle\strut q^{2}+m_{t}^{2}\deriv} & 0 \\
           0 & \frac{\displaystyle
        1\strut}{\displaystyle\strut q^{2}+m_{b}^{2}\deriv}
    \end{array}
    \right)
 \label{delta}
\ee
First, we calculate the trace of the sum
of all the products of $\gamma$-matrices displayed in
the expression (\ref{integral}). After that we come to the following
expression:
\ba
        &&\frac{N_{c}}{2}
        \int\limits_{|k|<\Lambda}\!\frac{d^4\!k}{(2\pi)^4}
        \Biggl\{2\,\tr\left[\mcal\dplus\mdag\dminus\right]
                \left(k^2-\frac{p^2}{4}\right)- \nonumber \\
        &&\quad-2\,\tr\left[\mdag \mavr\dplus\mdag \mavr\dminus\right]
                \label{integral2}+\\
        &&\quad+2\,\tr\left[\mdag\dplus\mcal\dminus\right]
                \left(k^2-\frac{p^2}{4}\right)- \nonumber \\
        &&\quad-2\,\tr\left[\mcal \mavr^{\dagger}\dplus\mcal
                \mavr^{\dagger}\dminus\right]
        \Biggr\}                \nonumber
\ea
In this formula and further on
the trace is applied only for the flavour two-by-two matrices.

The kinetic term is derived as being proportional to $p^2$ in
the soft-momentum expansion of (\ref{integral2}) in powers of $p^2$. We obtain this term by
means of calculating the second derivative of (\ref{integral2}) at zero
external momentum $ p $. First, let us rewrite the expression (\ref{integral2})
in a form:
\be
        \frac{N_{c}}{32\pi^4}\int\limits_{|k|<\Lambda}
        \!\!\left(f(x,y)-
        p^2\tr\left[\leftvertex\Delta(k^{2})
        \rightvertex\Delta(k^{2})\right]
        \right)\,d^{4}\!k,
\ee
where
\be
 x=\left(k+\frac{p}{2}\right)^{2},\qquad
 y=\left(k-{p\over 2}\right)^{2},       \label{xydefinition}
\ee
with the function
$ f $  defined as:
\ba
        f(x,y)&=&4 \tr\left[\leftvertex\Delta(x)
        \rightvertex\Delta(y)\right]k^{2}-\nonumber \\
        &-&2 \tr\left[\leftvertex \mavr^{\dagger}\Delta(x)
        \leftvertex \mavr^{\dagger}\Delta(y) \right]-\nonumber \\
        &-&2 \tr\left[\rightvertex \mavr\Delta(x)\rightvertex \mavr\Delta(y) \right]
        \label{integral3}
\ea
Let us expand the expression  (\ref{integral3}) in series of
$ p $ and extract the term proportional to
$ p^{2} $:
\be
        \frac{N_{c}p^{2}}{32\pi^2}\int\limits_{0}^{\Lambda^{2}}
                \left\langle\frac12
                \frac{\partial^{2}}{\partial p_{0}\partial p_{0}}\Biggr|_{p=0}
                f(x,y)-
                \tr\left[\leftvertex\Delta(k^{2})
                \rightvertex\Delta(k^{2})\right]
                \right\rangle k^{2}\,dk^{2},
\ee
where the angular brackets stand for angular average in 4-dimensional
Euclidean space and $p_0$ is a component of 4-momentum $p$.
The second derivative of the function
$ f $ reads:
\ba
        &&\left\langle
         \frac{\partial^{2}}{\partial p_{0}\partial p_{0}}\Biggr|_{p=0}
        f(x,y)
        \right\rangle = \no
        &&\quad=\quad\frac12\biggl[\left(f_{xx}(x_{0},y_{0})-
        2f_{xy}(x_{0},y_{0})+f_{yy}(x_{0},y_{0})
        \right)\frac{k^{2}}{2}\nonumber \\
        &&\quad+f_{x}(x_{0},y_{0})+
        f_{y}(x_{0},y_{0})
        \biggr]  \label{diffprocedure}
\ea
The subscripts in
$ f_{xx}, f_{xy}, f_{yy}, f_{x}, f_{y} $ stand for partial derivatives
by variables $ x $ and $ y $. The  derivatives are calculated at
$ x_{0}=y_{0}=k^{2} $ (see (\ref{xydefinition})).
The function
$ \Delta $, defined in (\ref{delta}), is a flavour matrix:
\be
    \Delta(x)=\left(
    \begin{array}{cc}
           \frac{
                \displaystyle
                1\strut}{
                \displaystyle
                \strut x+m_{t}^{2}\deriv} & 0 \\
           0 & \frac{
                \displaystyle
                1\strut}{
                \displaystyle
                \strut x+m_{b}^{2}\deriv}
    \end{array}    \label{prop2}
    \right).
\ee
Using (\ref{mmatrix}), (\ref{mmatrix2}), (\ref{mmatrix3}),
(\ref{integral3}) and (\ref{prop2}),
one gets
$ f(x,y) $:
\ba
        f(x,y)&=&\sum_{l,m=1}^{2}\Biggl\{\Biggl[
        4\Biggl(
          \frac{\fitl\fitm}{\xmt\ymb}+\nonumber \\
        &+&
          \frac{\fibl\fibm}{\xmb\ymt}
        \Biggr)k^{2}+\nonumber \\
        & +&2m_{t}\deriv m_{b}\deriv\Biggl(\fitl\fibm\faza{-}+\nonumber\\
        &+&
        \fibl\fitm\faza{}\Biggr)\times \nonumber \\
        &\times&\Biggl(\frac{1}{\xmt\ymb}+\nonumber \\
        &+&\frac{1}{\xmb\ymt}\Biggr)
        \Biggr]\philone^{\star}\phimone+ \nonumber \\
        &+&4k^{2}\Biggl[  \frac{\fibl\fibm}{\xmb\ymb}+\nonumber \\
        &+&
                \frac{\fitl\fitm}{\xmt\ymt}
        \Biggr]\,\philtwo^{\star}\phimtwo- \nonumber \\
        &-&2\Biggr[
        \frac{\fibl\fibm m_{b}^{2}\deriv\faza{-2}}{\xmb\ymb}+\nonumber \\
        &+&
                \frac{\fitl\fitm m_{t}^{2}\deriv}{\xmt\ymt}
        \Biggr]\philtwo\phimtwo- \nonumber \\
        &-&2\Biggr[
        \frac{\fibl\fibm m_{b}^{2}\deriv\faza{2}}{\xmb\ymb}+\nonumber \\
        &+&
                \frac{\fitl\fitm m_{t}^{2}\deriv}{\xmt\ymt}\Biggr]
        \philtwo^{\star}\phimtwo^{\star}
        \Biggr\}
\ea
After applying the derivative procedure displayed in (\ref{diffprocedure})
to the function
$ f $, one gets the kinetic term in the following form:
\ba
        &&\frac{N_{c}}{16\pi^2}
        \sum_{l,m=1}^{2}
        \Biggl(I_{lm}^{(1)}(\dmu \philone^{\star})(\dmu \phimone)+
        I_{lm}^{(2)}(\dmu \philtwo^{\star})(\dmu \phimtwo)+
        \nonumber \\
        &&\quad+
        I_{lm}^{(3)}(\dmu \philtwo)(\dmu \phimtwo)+
        I_{lm}^{(4)}
        (\dmu \philtwo^{\star})(\dmu \phimtwo^{\star})\Biggr). \label{sum1}
\ea
$ I^{(1)}_{lm} $ contributes to the kinetic term for charged components
of higgs doublets,
$ I^{(2)}_{lm} $,
$ I^{(3)}_{lm} $ and
$ I^{(4)}_{lm} $
do the same for the neutral components. The expressions
for them are cumbersome and  we have divided the total expression
in three parts.
For the charged components one has:
\ba
I_{lm}^{(1)}&=&\frac12 \int\limits_{0}^{\Lambda^{2}}\Biggl[
        \Biggl(
            \frac{\mtk{2} k^{2}}{\kmt^{3}\kmb}+\nonumber \\
        &+&
            \frac{\mbk{2}k^{2}}{\kmt\kmb^{3}}+\nonumber \\
        &+&
                \frac{k^{4}}{\kmt^{2}\kmb^{2}}+\nonumber \\
        &+&
                \frac{1}{\kmt\kmb}
        \Biggr)\times\nonumber \\
        &\times&
        \Biggl(\fitl\fitm+\fibl\fibm\Biggr)+\nonumber \\
        &+&
        \Biggl(\frac{\mtk{3}\mbk{}}{\kmt^{3}\kmb}+\nonumber \\
        &+&
                \frac{\mbk{3}\mtk{}}{\kmt\kmb^{3}}+\nonumber \\
        &+&
                \frac{\mtk{}\mbk{} k^{2}}{\kmt^{2}\kmb^{2}}
        \Biggr)\times\nonumber \\
        &\times&\Biggl(\fitl\fibm\faza{-}+\nonumber \\
        &+& \fibl\fitm\faza{}\Biggr)
\Biggr]\mesure,
\ea
for the neutral ones:
\ba
I_{lm}^{(2)}&=&\frac12 \int\limits_{0}^{\Lambda^{2}} \Biggl[
        \fitl\fitm\Biggl(
              -\frac{k^{4}}{\kmt^{4}}+\nonumber \\
        &+&
                \frac{2k^{2}}{\kmt^{3}}+
                \frac{1}{\kmt^{2}}
                \Biggr)+\nonumber \\
        &+&
        \fibl\fibm\Biggl(
              -\frac{k^{4}}{\kmb^{4}}+\nonumber \\
        &+&
                \frac{2k^{2}}{\kmb^{3}}+
                \frac{1}{\kmb^{2}}
                \Biggr)
\Biggr]\mesure
\ea
 and
\ba
I_{lm}^{(3)}&=&\frac12 \int\limits_{0}^{\Lambda^{2}} \Biggl[
        \fitl\fitm\left(
              \frac{k^{2}\mtk{2}}{2\kmt^{4}}-
                \frac{\mtk{2}}{\kmt^{3}}
                \right)+\nonumber \\
        &+&
        \fibl\fibm\faza{-2} \Biggl(
              -\frac{k^{2}\mbk{2}}{2\kmb^{4}}-\nonumber \\
        &-&
                \frac{\mbk{2}}{\kmb^{3}}
                \Biggr)
\Biggr]\!\mesure,
\ea
\ba
I_{lm}^{(4)}&=&\frac12 \int\limits_{0}^{\Lambda^{2}} \Biggl[
        \fitl\fitm\left(
              \frac{k^{2}\mtk{2}}{2\kmt^{4}}-
                \frac{\mtk{2}}{\kmt^{3}}
                \right)+\nonumber \\
        &+&
        \fibl\fibm\faza{2} \Biggl(
              -\frac{k^{2}\mbk{2}}{2\kmb^{4}}-\nonumber \\
        &-&
                \frac{\mbk{2}}{\kmb^{3}}
                \Biggr).
\Biggr]\!\mesure
\ea
The next task to do is to calculate the integrals for the large value
of
$ \Lambda $, ignoring all contributions which disappear in the
$ \Lambda\rightarrow\infty $ limit.
Thus one obtains:
\ba
        I_{lm}^{(1)}&=&\left(
           \fitlnull\fitmnull+\fiblnull\fibmnull
        \right)\times\nonumber \\
        &\times&\left[ \lgt
        +\frac{3\mb{4}\mt{2}-\mb{6}}{(\mt{2}-\mb{2})^{3}}\lgtb
                -\frac{\mt{4}+6\mt{2}\mb{2}+\mb{4}}{4(\mt{2}-\mb{2})^{2}}
        \right]+\nonumber \\
        &+&\left(\fitlnull\fibmnull\faza{-}
        +\fitlnull\fibmnull\faza{}\right)\times\nonumber \\
        &\times&\left[ -\frac{2\mt{3}\mb{3}}{(\mt{2}-\mb{2})^{3}}\lgtb
        +\frac{\mt{}\mb{}(\mt{2}+\mb{2})}{4(\mt{2}-\mb{2})^{2}}
        \right]+\nonumber \\
        &+&\int\limits_{0}^{1}\bigl(\fitltau\fitmtau
                +\fibltau\fibmtau-\nonumber \\
       &&-\fitlnull\fitmnull-\fiblnull\fibmnull
        \bigr)\mesuretwo+\remnant,
\ea
\ba
        I_{lm}^{(2)}&=&\fitlnull\fitmnull\left[
           \lgt-\frac{13}{12}
        \right]
        +\fiblnull\fibmnull\left[
           \lgb-\frac{13}{12}
        \right]+\nonumber \\
        &+&\int\limits_{0}^{1}\Biggl(\fitltau\fitmtau
                +\fibltau\fibmtau-\nonumber \\
        &&-\fitlnull\fitmnull-\fiblnull\fibmnull
        \Biggr)\mesuretwo+\remnant,
\ea
\be
        I_{kl}^{(3)}=
-\frac13 (\fitlnull\fitmnull+\fiblnull\fibmnull\faza{-2})+\remnant,
\ee
\be
        I_{kl}^{(4)}=
-\frac13 (\fitlnull\fitmnull+\fiblnull\fibmnull\faza{2})+\remnant.
\ee
Herein and further on
$ m_{t} $ and
$ m_{b} $ stand for quark masses.

As we know from experiment, the mass of the top quark is much greater than
that of the bottom quark. Regarding $\mt{}\gg\mb{}$, for the choice of
formfactors (\ref{formfactors})  one gets:
\ba
        I_{11}^{(1)}&\approx&\lgt-\frac14 + 2\ctone+\frac12 (\ctone^2+\cbone^2),\\
        I_{12}^{(1)}&\approx&\cttwo+\cbone+\frac12 (\ctone\cttwo+\cbone\cbtwo),\\
        I_{22}^{(1)}&\approx&\lgb-\frac14 + 2\cbtwo+\frac12 (\cttwo^2+\cbtwo^2),
\ea

\ba
        I_{11}^{(2)}&\approx&\lgt-\frac{13}{12} +
                2\ctone+\frac12 (\ctone^2+\cbone^2),\\
        I_{12}^{(2)}&\approx&\cttwo+\cbone+\frac12 (\ctone\cttwo+\cbone\cbtwo),\\
        I_{22}^{(2)}&\approx&\lgb-\frac{13}{12} + 2\cbtwo+
                \frac12 (\cttwo^2+\cbtwo^2),
\ea

\ba
        I_{11}^{(3)}&\approx&-\frac13, \\
        I_{12}^{(3)}&\approx&0, \\
        I_{22}^{(3)}&\approx&-\frac13\faza{-2}.
\ea
\ba
        I_{11}^{(4)}&\approx&-\frac13, \\
        I_{12}^{(4)}&\approx&0, \\
        I_{22}^{(4)}&\approx&-\frac13\faza{2}
\ea
Next, we shall change variables and rewrite the total expression for
kinetic term. For  the neutral components we choose  non-linear
parametrization:
\be
        \phi_1e^{i\alpha}\equiv \phi_{12}, \qquad \phi_2 e^{i(\alpha+\delta)}\equiv
        \phi_{22}. \label{newvariables}
\ee
Let us substitute (\ref{newvariables}) into (\ref{sum1}) and expand it with
derivatives, leaving only quadra\-tic terms and omitting the rest of the
expression.
As the v.e.v. of
$ \phi_{1} $
and
$ \phi_{2} $ provide quarks with dynamical masses, we replace them
with
$ m_{t} $ and
$ m_{b} $ respectively.
The variable
$\alpha$  is regarded as a Goldstone boson, which is absent in
the  effective potential; it appears only in higher-derivative  terms.
The other phase, $\delta$ , is associated with the relative phase.
The variables
$\phi_1$ and $\phi_2$ parameterize radial excitations.
For the fields $\alpha$
and $\delta$ we use different notations
\be
     \delta  \equiv  \phi_3,\qquad \alpha  \equiv  \phi_4 , \label{newdef}
\ee
so that one can rewrite the kinetic term for the neutral components of
Higgs doublets uniformly:
\be
       \frac{N_c}{16\pi^2} \sum_{l,m=1}^{4} A_{lm}(\dmu\phi_{l})
                (\dmu\phi_m),
\ee
where $A$ is four-by-four matrix:
\be
        \left(
        \begin{array}{l|l|l|l}
        \displaystyle\strut
        I_{11}^{(2)}-\frac23
        \displaystyle\strut
                & I_{12}^{(2)}\cos\delta_{0}
        \displaystyle\strut
                    & -I_{12}^{(2)}m_b\sin\delta_{0}
        \displaystyle\strut
                        & -I_{12}^{(2)}m_b\sin\delta_{0}\\ \hline
        \displaystyle\strut
           I_{12}^{(2)}\cos\delta_{0}
                & \displaystyle\strut
                 I_{22}^{(2)}-\frac23
                    & 0
                         &\displaystyle\strut
                         I_{12}^{(2)}m_t\sin\delta_{0}\\ \hline
        \displaystyle\strut
          -I_{12}^{(2)}m_b\sin\delta_{0}
                & 0
                    &    \displaystyle\strut
                     (I_{12}^{(2)}+\frac23)\mb{2}
                        & \displaystyle\strut
                         (I_{22}^{(2)}+\frac23)\mb{2}+\\
        &&&
        \displaystyle\strut
                         +I_{12}^{(2)}\mb{}\mt{}\cos\delta_{0}
        \\ \hline
        \displaystyle\strut
          -I_{12}^{(2)}m_{b}\sin\delta_{0}
                & \displaystyle\strut
                  I_{12}^{(2)}m_t\sin\delta_{0}
                    & \displaystyle\strut
                     I_{12}^{(2)}\mb{}\mt{}\cos\delta_{0}+
                        & \displaystyle\strut
                         (I_{11}^{(2)}+\frac23)\mt{2}+
                         \\
           &&
        \displaystyle\strut
                        +(I_{22}^{(2)}+\frac23)\mb{2}
           &
        \displaystyle\strut
                +2I_{12}^{(2)}\mt{}\mb{}\cos\delta_{0}+\\
        &&&   +(I_{22}^{(2)}+\frac23)\mb{2}
        \end{array}
        \right).
\ee

\section*{Appendix B:  Momentum independent matrix $\hat {B}$}

Let us define the matrix of second variations of the effective potential
for the  Model I in  the following way:
\be
B_{kl}^{\pi\sigma}=\frac{8\pi^2}{N_{c}}\frac{\partial^{2}}{\partial\!\phi_{l}
\partial\!\phi_{m}}V_{eff} ,\qquad (l,m=1,2).
\ee
For the case, when $\rho =0$,
the $ B_{kl}^{\sigma\sigma} $ matrix for scalars and $ B_{kl}^{\pi\pi} $  matrix
for  pseudoscalars are represented:
\begin{eqnarray}
B_{11}^{\sigma\sigma}&=&
        (1+g^{4})\left[
        64\phi_{1}^{2}
        \ln\left({{\Lambda^{2}}\over{4\phi_{1}^{2}}}\right)
        -223\phi_{1}^{2}-\frac{15\sqrt{3}}{2}\phi_{1}\phi_{2}-
        \frac{\sqrt{3}}{2}\frac{\phi_{2}^{3}}{\phi_{1}}
        \right]+\nonumber\\
        &&+2\Delta_{12}\frac{\phi_{2}}{\phi_{1}}
        -64\phi_{1}^{2}g^{4}\ln\,g^2,\\
B_{12}^{\sigma\sigma}&=&
       -2\Delta_{12}+(1+g^{4})\left[
        9\phi_{1}\phi_{2}-\frac{15\sqrt{3}}{2}\phi_{1}^{2}+
        \frac{3\sqrt{3}}{2}\phi_{2}^{2}
        \right],   \\
B_{22}^{\sigma\sigma}&=&
        (1+g^{4})\left[9\phi_{2}^{2}+
        \frac{3\sqrt{3}}{2}\phi_{1}\phi_{2}+
        \frac{5\sqrt{3}}{2}\frac{\phi_{1}^{3}}{\phi_{2}}\right]+
        2\Delta_{12}\frac{\phi_{1}}{\phi_{2}},\\
B_{11}^{\pi\pi}&=&
        (1+g^{4})\left[
        \frac{5\sqrt{3}}{2}\phi_{1}\phi_{2}-
        \frac{\sqrt{3}}{2}\frac{\phi_{2}^{3}}{\phi_{1}}-
        3\phi_{2}^{2}
        \right]+2\Delta_{12}\frac{\phi_{2}}{\phi_{1}},\\
B_{12}^{\pi\pi}&=&
        -2\Delta_{12}+(1+g^{4})
        \left[
        \frac{5\sqrt{3}}{2}\phi_{1}^{2}+\frac{\sqrt{3}}{2}\phi_{2}^{2}
        +3\phi_{1}\phi_{2}
        \right], \\
B_{22}^{\pi\pi} &=&
        (1+g^{4})\left[
        -3\phi_{1}^{2}-\frac{\sqrt{3}}{2}\phi_{1}\phi_{2}+
        \frac{5\sqrt{3}}{2}\frac{\phi_{1}^{3}}{\phi_{2}}
        \right]+2\Delta_{12}\frac{\phi_{1}}{\phi_{2}},\\
B_{kl}^{\sigma\pi}&=&0\qquad k,l=(1,2).
\end{eqnarray}
For the case, $\rho $ non-zero,
the corresponding matrix of second variations of the effective potental
is a
$ 4\times 4 $ matrix. One can arrange the neutral Higgses in
a vector-column with 4 components:
$$
  \left(
        \begin{array}{c}
        \sigma_{1}\\
        \sigma_{2}\\
        \pi_1 \\
        \pi_2
        \end{array}
\right),
$$
where
$ \sigma_{k} $ - scalar fields, $ \pi_{k} $ - pseudoscalar.

Let us display the non-vanishing components of the matrix of
second variations, calculated in the minimum of the potential,
in the following form:
 \begin{eqnarray}
B_{11}^{\sigma\sigma}&=&
        (1+g^{4})\left[
        64\phi_{1}^{2}
        \ln\left({{\Lambda^{2}}\over{4\phi_{1}^{2}}}\right)
        -223\phi_{1}^{2}-10\sqrt{3}\phi_{1}\phi_{2}+3\phi_{2}^{2}
        \right]-\nonumber\\
        &&-64\phi_{1}^{2}g^{4}\ln\,g^2,\\
B_{12}^{\sigma\sigma}&=&
       (1+g^{4})\left[-5\sqrt{3}\phi_{1}^{2}+
        6\phi_{1}\phi_{2}+
        \sqrt{3}\phi_{2}^{2}
        \right],   \\
B_{22}^{\sigma\sigma}&=&
        (1+g^{4})\left[
        3\phi_{1}^{2}+
        2\sqrt{3}\phi_{1}\phi_{2}+
        9\phi_{2}^{2}
        \right],\\
B_{11}^{\pi\pi}&=&
        3(1+g^{4})\rho^{2},\\
B_{12}^{\pi\pi}&=&
        \sqrt{3}(1+g^{4})\rho^{2},\\
B_{22}^{\pi\pi} &=&
        9(1+g^{4})\rho^{2},\\
B_{11}^{\sigma\pi}&=&
        \rho(1+g^{4})\left(-5\sqrt{3}\phi_{1}+3\phi_{2}\right),\\
B_{12}^{\sigma\pi}&=&
        \rho(1+g^{4})\left(3\phi_{1}+\sqrt{3}\phi_{2}\right),\\
B_{21}^{\sigma\pi}&=&
        \rho(1+g^{4})\left(3\phi_{1}+\sqrt{3}\phi_{2}\right),\\
B_{22}^{\sigma\pi}&=&
        \rho(1+g^{4})\left(\sqrt{3}\phi_{1}+9\phi_{2}\right).
\end{eqnarray}

\bigskip
%(the common factor ${N_{c}/8\pi^2}$ is implied).

The matrix $\hat {B}$ for the effective potential in the
Model II is:
\ba
B_{11}&=&
-2\Delta_{11}+6\phi_{1}^{2}\ln\left(\frac{\Lambda^{2}}{\phi_{1}^{2}}\right)-
        4\phi_{1}^{2}+6J_{1111}\phi_{1}^{2}+\nonumber \\
        &&+12J_{1112}\phi_{1}\phi_{2}\cos\delta_{0}+
        2\left(
        J_{1122}+J_{1221}+J_{1212}\cos2\delta_{0}
        \right)\phi_{2}^{2},\label{bmatrixbegin}\\
B_{12}&=&
-2\Delta_{12}\cos\delta_{0}+6J_{1112}\phi_{1}^{2}\cos\delta_{0}
        + 6J_{1222}\phi_{2}^{2}\cos\delta_{0}+\nonumber\\
        &&+4\left(
        J_{1122}+J_{1221}+J_{1212}\cos2\delta_{0}
        \right)\phi_{1}\phi_{2},\\
B_{13}&=&
        \phi_{2}\sin\delta_{0}
        \left(2\Delta_{12}-6J_{1112}\phi_{1}^{2}-
        8J_{1212}\phi_{1}\phi_{2}\cos\delta_{0}-
        2J_{1222}\phi_{2}^{2}
        \right),\\
B_{14}&=&0,\\
B_{21}&=&B_{12},\\
B_{22}&=&
-2\Delta_{22}+6\phi_{2}^{2}\ln\left(\frac{\Lambda^{2}}{\phi_{2}^{2}}\right)-
        4\phi_{2}^{2}+12J_{1222}\phi_{1}\phi_{2}\cos\delta_{0}+\nonumber \\
        &&+6J_{2222}\phi_{2}^{2}+
        2\left(
        J_{1122}+J_{1221}+J_{1212}\cos2\delta_{0}
        \right)\phi_{1}^{2},\\
B_{23}&=&
        \phi_{1}\sin\delta_{0}
        \left(2\Delta_{12}-6J_{1222}\phi_{2}^{2}-
        8J_{1212}\phi_{1}\phi_{2}\cos\delta_{0}-
        2J_{1112}\phi_{1}^{2}
        \right),\\
B_{24}&=&0,\\
B_{31}&=&B_{13}\\
B_{32}&=&B_{23},\\
B_{33}&=&
        2\phi_{1}\phi_{2}\bigl(
        \Delta_{12}\cos\delta_{0}-
        J_{1112}\phi_{1}^{2}\cos\delta_{0}-
        2J_{1212}\phi_{1}\phi_{2}\cos2\delta_{0}-\nonumber\\
        &&-
        J_{1222}\phi_{2}^{2}\cos\delta_{0}        \bigr),\\
B_{41}&=&0,\\
B_{42}&=&0,\\
B_{43}&=&0,\\
B_{44}&=&0 .\label{bmatrixend}
\ea

\section*{Appendix C: The effective potential of Model II
for the special choice of formfactors}

For purposes of further calculations of realistic mass spectra, Yukawa coupling
constants and decay widths with taking into account the renormalization-group
corrections we present in this appendix the effective
potential of Model II for the following set of formfactors:
\ba
  &&f_{t,1}=2-3\tau; \qquad f_{t,2}=-\sqrt{3}\tau;\nonumber\\
  &&f_{b,1}=-\sqrt{3}\tau; \qquad  f_{b,2}=2-3\tau;
\ea
the constants
$ J_{klmn} $ are evaluated to definite numbers. Seven of then are
defined as follows:
\begin{eqnarray}
J_{1111}&=&\frac34\left(3g_{b,1}^{4}-53g_{t,1}^{4}\right)\label{J1},\\
J_{1112}&=&\frac{\sqrt{3}}{4}\left(g_{b,1}^{3}g_{b,2}-5g_{t,1}^{3}g_{t,2}\right),\\
J_{1122}&=&\frac34\left(g_{b,1}^{2}g_{b,2}^{2}-53 g_{b,2}^{2}g_{t,1}^{2}
 -2g_{b,1}g_{b,2}g_{t,1}g_{t,2}+
  3g_{b,1}^{2}g_{t,2}^{2}+g_{t,1}^{2}g_{t,2}^{2}
  \right),\\
J_{1212}&=&\frac34\left(g_{b,1}^{2}g_{b,2}^{2}+g_{t,1}^{2}g_{t,2}^{2}\right),\\
J_{1221}&=&\frac34\left(g_{b,1}^{2}g_{b,2}^{2}+53 g_{b,2}^{2}g_{t,1}^{2}
 +2g_{b,1}g_{b,2}g_{t,1}g_{t,2}-
  3g_{b,1}^{2}g_{t,2}^{2}+g_{t,1}^{2}g_{t,2}^{2}
   \right),\\
J_{1222}&=&\frac{\sqrt{3}}{4}\left(g_{t,1}g_{t,2}^{3}-5g_{b,1}g_{b,2}^{3}\right),\\
J_{2222}&=&\frac34\left(3g_{t,2}^{4}-53g_{b,2}^{4}\right).\label{J2}
\end{eqnarray}
The rest of them is found from their symmetry property:
$$
  J_{klmn}=J_{mnkl}=J_{lknm}.
$$
With (\ref{J1})--(\ref{J2}) the potential for the Model II reads:
\begin{eqnarray}
V_{eff}&=&
{{N_{c}}\over{8\pi^2}}\Biggl\{
  -\Delta_{11}(\Phi_{1}^{\dagger}\Phi_{1})
  -\Delta_{12}(\Phi_{1}^{\dagger}\Phi_{2})
  -\Delta_{21}(\Phi_{2}^{\dagger}\Phi_{1})
  -\Delta_{22}(\Phi_{2}^{\dagger}\Phi_{2})+\nonumber\\
&&+\frac{\nu_{+}^{2}(0)}{2}\left(\ln{{\Lambda^{2}}\over{\nu_{+}(0)}}+\frac12\right)+
\frac{\nu_{-}^{2}(0)}{2}\left(\ln{{\Lambda^{2}}\over{\nu_{-}(0)}}+\frac12\right)+
\nonumber\\
&&+\frac38\left(3g_{b,1}^{4}-53g_{t,1}^{4}\right)(\Phi_{1}^{\dagger}\Phi_{1})^{2}+
\frac38\left(3g_{t,2}^{4}-53g_{b,2}^{4}\right)(\Phi_{2}^{\dagger}\Phi_{2})^{2}+
\nonumber\\
&&+\frac34\left(g_{b,1}^{2}g_{b,2}^{2}-53 g_{b,2}^{2}g_{t,1}^{2}
 -2g_{b,1}g_{b,2}g_{t,1}g_{t,2}+
  3g_{b,1}^{2}g_{t,2}^{2}+g_{t,1}^{2}g_{t,2}^{2}
  \right)(\Phi_{1}^{\dagger}\Phi_{1})(\Phi_{2}^{\dagger}\Phi_{2})+
\nonumber\\
&&+\frac34\left(g_{b,1}^{2}g_{b,2}^{2}+53 g_{b,2}^{2}g_{t,1}^{2}
 +2g_{b,1}g_{b,2}g_{t,1}g_{t,2}-
  3g_{b,1}^{2}g_{t,2}^{2}+g_{t,1}^{2}g_{t,2}^{2}
   \right)(\Phi_{1}^{\dagger}\Phi_{2})(\Phi_{2}^{\dagger}\Phi_{1})+
\nonumber\\
&&+\frac38\left(g_{b,1}^{2}g_{b,2}^{2}+g_{t,1}^{2}g_{t,2}^{2}\right)
\left[(\Phi_{1}^{\dagger}\Phi_{2})^{2}+
(\Phi_{1}^{\dagger}\Phi_{2})^{2}\right]+
\nonumber\\
&&+\frac{\sqrt{3}}{4}\left(g_{b,1}^{3}g_{b,2}-5g_{t,1}^{3}g_{t,2}\right)
 (\Phi_{1}^{\dagger}\Phi_{1})\left[
 (\Phi_{1}^{\dagger}\Phi_{2})+ (\Phi_{2}^{\dagger}\Phi_{1})
\right]+
\nonumber\\
&&+\frac{\sqrt{3}}{4}\left(g_{t,1}g_{t,2}^{3}-5g_{b,1}g_{b,2}^{3}\right)
 (\Phi_{2}^{\dagger}\Phi_{2})\left[
 (\Phi_{1}^{\dagger}\Phi_{2})+ (\Phi_{2}^{\dagger}\Phi_{1})
\right]
\Biggr\}
\end{eqnarray}
where we adopt the definition for $\nu_{\pm}(0)$:
\begin{eqnarray}
&&\nu_{\pm}=
2g_{t,1}^{2}(\Phi_{1}^{\dagger}\Phi_{1})+
2g_{b,2}^{2}(\Phi_{2}^{\dagger}\Phi_{2})\pm
2\Bigl[
g_{t,1}^{4}(\Phi_{1}^{\dagger}\Phi_{1})^{2}+
g_{b,2}^{4}(\Phi_{2}^{\dagger}\Phi_{2})^{2}+
\nonumber\\
&&\qquad+
2g_{t,1}^{2}g_{b,2}^{2}(\Phi_{1}^{\dagger}\Phi_{1})
(\Phi_{2}^{\dagger}\Phi_{2})-
4g_{t,1}^{2}g_{b,2}^{2}(\Phi_{1}^{\dagger}\Phi_{2})
(\Phi_{2}^{\dagger}\Phi_{1})
\Bigr]^{1/2}
\end{eqnarray}

\end{document}